\documentclass[aps,prd,groupedaddress,showpacs]{revtex4}
\usepackage{amsmath}
\usepackage{amssymb}
\usepackage{bbm}
\usepackage{epsfig}
\begin{document}
\title{Combined effects of strong and electroweak FCNC effective operators in
top quark physics at the LHC}

\author{P.M. Ferreira$^{1,2}$~\footnote{ferreira@cii.fc.ul.pt}, R.B.
Guedes$^{2}$~\footnote{renato@cii.fc.ul.pt} and R.
Santos$^{2,3}$~\footnote{rsantos@cii.fc.ul.pt}} \affiliation{$^1$
Instituto Superior de Engenharia de Lisboa, Rua Conselheiro Em\'{\i}dio
Navarro, 1, 1959-007 Lisboa, Portugal \\
$^{2}$ Centro de F\'{\i}sica Te\'orica e Computacional, Faculdade de
Ci\^encias, Universidade de Lisboa, Avenida Professor Gama Pinto, 2,
1649-003
Lisboa, Portugal \\
$^{3}$ Department of Physics, Royal Holloway, University of London,
Egham, Surrey TW20 0EX United Kingdom }

\date{February, 2008}

\begin{abstract}
We study the combined effects of both strong and electroweak
dimension six effective operators on flavour changing top quark
physics at the LHC. Analytic expressions for the cross sections and
decay widths of several flavour changing processes will be
presented, as well as an analysis of the feasibility of their
observation at the LHC.
\end{abstract}


\maketitle

\section{Introduction}

The LHC will soon begin operating, and the number of top quarks
produced in it is of the order of millions per year. Such large
statistics will enable precision studies in top quark physics - this
being the least well-know elementary particle discovered so far. The
study of flavour changing neutral current (FCNC) interactions of the
top quark is of particular interest. In fact, the FCNC decays of the
top - decays to a quark of a different flavour and a gauge boson, or
a Higgs scalar - have branching ratios which can vary immensely from
model to model - from the extremely small values expected within the
Standard Model (SM) to magnitudes possibly measurable at the LHC in
certain SM extensions.

The use of anomalous couplings to study possible new top physics at
the LHC and Tevatron has been the subject of many works~\cite{whis}.
In a recent series of papers~\cite{Ferr1,Ferr2,Ferr3} we considered
FCNC interactions associated with the strong interaction - decays of
the type $t\rightarrow u\,g$ or $t\rightarrow c\,g$ - describing
them using the most general dimension six FCNC lagrangian emerging
from the effective operator formalism~\cite{buch}. The FCNC vertices
originating from that lagrangian also had substantial contributions
to processes of production of the top quark, such as associated
production of a single top quark alongside a jet, a Higgs boson - or
an electroweak gauge boson. The study of
refs.~\cite{Ferr1,Ferr2,Ferr3} concluded that, for large values of
$BR(t\rightarrow q\,g)$, with $q = u, c$, these processes of single
top production might be observable at the LHC.

What about the possibility of FCNC associated with the electroweak
sector - FCNC interactions leading to decays of the form
$t\rightarrow q\,\gamma$ or $t\rightarrow q\,Z$ ? In some extensions
of the SM these branching ratios can be as large as, if not larger,
those of the strong FCNC interactions involving gluons. In the
current paper we extend the analysis of our previous works and
consider the most general dimension six FCNC lagrangian in the
effective operator formalism which leads to $t\rightarrow q\,\gamma$
and $t\rightarrow q\,Z$ decays. We will study the effects of these
new electroweak FCNC interactions in the decays of the top quark and
its expected production at the LHC. We will study in detail
processes - such as $t\,+\,\gamma$ and $t\,+\,Z$ production - for
which both strong and electroweak FCNC interactions contribute. The
automatic gauge invariance of the effective operator formalism will
allow us to detect correlations between several FCNC observables.
The FCNC processes $pp \rightarrow t \, Z$ and $pp \rightarrow t \,
\gamma$ were studied in great detail for the Tevatron
in~\cite{delAguila:1999ac} and for the LHC in~\cite{del
Aguila:1999ec}. We will draw heavily on the results of those
references, all the while emphasising the differences in our
approaches: (a) our chief aim is to provide the scientific community
with analytical expressions anyone can use to built event generators
and perform detailed studies of FCNC at the LHC; (b) we show all
results in terms of measurable quantities, such as branching ratios,
and not in terms of the values of the anomalous couplings; and (c),
our formalism leads us to to write FCNC vertices different from
those of refs.~\cite{delAguila:1999ac, del Aguila:1999ec}, and to
uncover connections between several FCNC quantities.

This paper is organised as follows: in section~\ref{sec:eff} we
review the effective operator formalism and introduce our FCNC
operators, explaining what physical criteria were behind their
choice. We also present the Feynman rules for the new anomalous top
quark interactions which will be the base of all the work that
follows. In section~\ref{sec:brs} we use those same Feynman rules to
compute and analyse the branching ratios of the top quark FCNC
decays, with particular emphasis on the relationship between
$Br(t\,\rightarrow\,q\,\gamma)$ and $Br(t\,\rightarrow\,q\,Z)$. In
the following two sections we study the cross section for
production, at the LHC, of a single top and a photon or a $Z$ boson,
with all FCNC interactions - both strong and electroweak - included.
We also investigate whether it would be possible to conclude, from
the data, whether any FCNC phenomena observed would have at its root
the strong or the electroweak sectors. Finally, in
section~\ref{sec:disc} we present a general discussion of the
results and some conclusions.

\section{Flavour changing effective operators}
\label{sec:eff}

\noindent The effective operator formalism of Buchm\"uller and
Wyler~\cite{buch} is based on the assumption that the Standard Model
of particle physics is the low energy limit of a more general
theory. Such theory would be valid at very high energies but, at a
lower energy scale $\Lambda$, we would only perceive its effects
through a set of effective operators of dimensions higher than four.
Those operators would obey the gauge symmetries of the SM, and be
suppressed by powers of $\Lambda$. This allows us to write this
effective lagrangian as a series, such that 
\begin{equation}
{\cal L} \;\;=\;\; {\cal L}^{SM} \;+\; \frac{1}{\Lambda}\,{\cal
L}^{(5)} \;+\; \frac{1}{\Lambda^2}\,{\cal L}^{(6)} \;+\;
O\,\left(\frac{1}{\Lambda^3}\right) \;\;\; , \label{eq:l}
\end{equation}
where ${\cal L}^{SM}$ is the SM lagrangian and ${\cal L}^{(5)}$ and
${\cal L}^{(6)}$ contain all the dimension five and six operators
which, like ${\cal L}^{SM}$, are invariant under the gauge
symmetries of the SM. The list of dimension six operators is quite
vast~\cite{buch}. This formalism allows us to parameterize new
physics, beyond that of the SM, in a model-independent manner.

In this work we are interested in effective operators of dimension
six that contribute to flavour-changing interactions of the top
quark in the weak sector. The ${\cal L}^{(5)}$ terms break baryon
and lepton number conservation, and therefore we do not consider
them in this analysis. This work follows
refs.~\cite{Ferr1,Ferr2,Ferr3}, where we considered FCNC top
effective operators which affect the strong sector. Namely,
operators which, amongst other things, contribute to FCNC decays of
the form $t\, \rightarrow \,u\,g$ or $t\, \rightarrow \,c\,g$. The
operators we considered were expressed as
\begin{align}
{\cal O}_{tG} &= i \frac{\alpha^S_{it}}{\Lambda^2}\,
\left(\bar{u}^i_R \, \lambda^{a} \, \gamma_{\mu}  D_{\nu} t_R\right)
\, G^{a \mu \nu} \;\;\;,\;\;\;{\cal
O}_{tG\phi}=\frac{\beta^{S}_{it}}{\Lambda^2}\,\left(\bar{q}^i_L\, \,
\lambda^{a} \, \sigma^{\mu\nu}\,t_R\right)\, \tilde{\phi}
\,G^{a}_{\mu\nu}\;\;\; , \label{eq:opst}
\end{align}
where the coefficients $\alpha^S_{it}$ and $\beta^{S}_{it}$ are
complex dimensionless couplings. The fields $u^i_R$ and $q^i_L$
represent the right-handed up-type quark and left-handed quark
doublet of the first and second generation - this way FCNC occurs.
$G^a_{\mu\nu}$ is the gluonic field tensor. There are also
operators, with couplings $\alpha^S_{ti}$ and $\beta^{S}_{ti}$,
where the positions of the top and $u^î$, $q^i$ spinors are
exchanged in the expressions above. Also, the hermitian conjugates
of all of these operators are obviously included in the lagrangian.
These operators contribute to FCNC vertices of the form
$g\,t\,\bar{u_i}$ (with $u_i \,=\,u\,,\,c$). The operators with
$\alpha^S$ couplings, due to their gauge structure (namely, the
covariant derivative acting on a quark spinor), also contribute to
quartic vertices of the form $g\,g\,t\,\bar{u_i}$,
$g\,\gamma\,t\,\bar{u_i}$ and $g\,Z\,t\,\bar{u_i}$.

Our criteria in choosing these operators were that they contributed
only to FCNC top physics, not affecting low energy physics. In that
sense, operators that contributed to top quark phenomenology but
which also affected bottom quark physics (in the notation of
ref.~\cite{buch}, operators ${\cal O}_{qG}$) were not considered.
Recently, a study based on constraints from B
physics~\cite{Fox:2007in} using the predictions for the
LHC~\cite{toni, fla, CMS}, has showed that, in fact, some of the
constraints on dimension 6 operators stemming from low energy
physics are already stronger than some of the predictions for the
LHC. This is true for the operators denoted in~\cite{Fox:2007in} by
$LL$, which are the ones built with two $SU(2)$ doublets that we had
left out in our previous work. Obviously the gauge structure is felt
more strongly in the left-left (LL) type of operators than in the
right-right type. Hence, they concluded that the LL operators will
not be probed at the LHC because they are already constrained beyond
the expected bounds obtained for a luminosity of 100 $fb^{-1}$.
Limits on LR and RL operators are close to those experimental bounds
and RR operators are the ones that will definitely be probed at the
LHC. Moreover, since more results will come from the B factories and
the Tevatron, the constraints will be even stronger by the time the
LHC starts to analyse data. Therefore our criteria in the choice of
operators is well founded, and we will also not consider LL
operators in the electroweak sector.

\subsection{Effective operators contributing to electroweak FCNC top decays}
\noindent
According to our criteria of leaving low-energy particle physics
unchanged, we will now consider all possible dimension six effective
operators which contribute to top decays of the form
$t\,\rightarrow\,u_i\,\gamma$ and $t\,\rightarrow\,u_i\,Z$. First we
have the operators analogous to those of eq.~\eqref{eq:opst} in the
electroweak sector, to wit,
\begin{align}
{\cal O}_{tB}= i \frac{\alpha^B_{it}}{\Lambda^2}\,\left(\bar{u}^i_R
\, \, \gamma_{\mu} D_{\nu} t_R \right) \, B^{\mu \nu}\;\;\; , &
\;\;\;{\cal O}_{tB\phi} =
\;\;\frac{\beta^{B}_{it}}{\Lambda^2}\,\left(\bar{q}^i_L\,
\sigma^{\mu\nu}\,t_R\right)\, \tilde{\phi} \,B_{\mu\nu} \;\;\; , \nonumber \\
{\cal
O}_{tW\phi}=\frac{\beta^{W}_{it}}{\Lambda^2}\,\left(\bar{q}^i_L\, \,
\tau_{I} \, \sigma^{\mu\nu}\,t_R\right)\,
\tilde{\phi}\,W^{I}_{\mu\nu}\;\;\; , & \label{eq:op4}
\end{align}
where $\alpha^B_{ti}$, $\beta^{B}_{ti}$ and $\beta^{W}_{ti}$  are
complex dimensionless couplings, and $B^{\mu \nu}$ and
$W^{I}_{\mu\nu}$ are the $U(1)_Y$ and $SU(2)_L$ field tensors,
respectively. As before, we also consider the operators with
exchanged quark spinors, corresponding to couplings $\alpha^B_{ti}$,
$\beta^{B}_{ti}$ and $\beta^{W}_{ti}$, and the hermitian conjugates
of all of these terms.

The electroweak tensors ``contain" both the photon and $Z$ boson
fields, through the well-known Weinberg rotation. Thus they
contribute simultaneously to vertices of the form $Z \,\bar{t} \,
u_i$ and $\gamma\, \bar{t} \, u_i$ when we consider the partial
derivative of $D^\mu$ in the equations~\eqref{eq:op4}, or when we
replace the Higgs field $\phi$ by its vev $v$ in them. We will
isolate the contributions to FCNC photon and $Z$ interactions in
these operators defining new effective couplings
$\{\alpha^\gamma\,,\,\beta^{\gamma}\}$ and
$\{\alpha^Z\,,\,\beta^Z\}$. These are related to the initial
couplings via the Weinberg angle $\theta_W$ by
\begin{equation}
\alpha^{\gamma}\;=\;\cos\theta_W \, \alpha^{B} \qquad \; , \; \qquad
\alpha^{Z}\;=\; - \sin\theta_W \, \alpha^{B} \label{eq:alf}
\end{equation}
and
\begin{equation}
\left\{
\begin{array}{c}
   \beta^{\gamma} \, = \, \sin\theta_W \beta^{W} + \cos \theta_W \beta^{B}\\
  \beta^{Z} \, = \, \cos\theta_W \beta^{W} - \sin \theta_W \beta^{B}  \\
\end{array}
\right. . \label{eq:bet}
\end{equation}
As we will see, these Weinberg rotations will introduce a certain
correlation between FCNC processes involving the photon or the $Z$.

Because the Higgs field is electrically neutral but has weak
interactions, there are more effective operators which will only
contribute to new $Z$ FCNC interactions. They are analogous to
operators considered in~\cite{Lept} for study of FCNC in the
leptonic sector and are given by
\begin{align}
{\cal O}_{D_t} &=\frac{\eta_{it}}{\Lambda^2}\,\left(\bar{q}^i_L\,
D^{\mu}\,t_R\right)\, D_{\mu} \tilde{\phi} \, \;\;\;,\;\;\; {\cal
O}_{\bar{D}_t}=\frac{\bar{\eta}_{it}}{\Lambda^2}\,\left( D^{\mu}
\bar{q}^i_L\, \,t_R\right)\, D_{\mu} \tilde{\phi}
\label{eq:op1}
\end{align}
and
\begin{equation}
{\cal O}_{\phi_t}
 \, = \, \theta_{it} \, (\phi^{\dagger} D_{\mu} \phi) \, (\bar{u^i_R} \gamma^{\mu} t_R)  \;\;\;
 ,
\label{eq:op2}
\end{equation}
and another operator with coupling $\theta_{ti}$ with the position
of the $u^i$ and $t$ spinors exchanged. As before, the coefficients
$\eta_{it}$, $\bar{\eta}_{it}$ and $\theta_{it}$ are complex
dimensionless couplings.

\subsection{Feynman rules for top FCNC weak interactions}

\begin{figure}[htbp]
  \begin{center}
    \epsfig{file=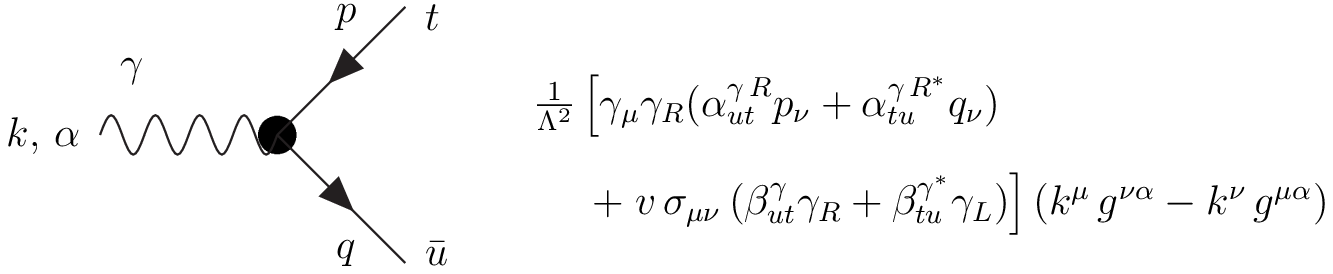,width=14 cm}
    \caption{Feynman rules for the anomalous vertex $\gamma \, t\, \bar{u} $.}
    \label{fig:feyngamma}
  \end{center}
\end{figure}

\begin{figure}[htbp]
  \begin{center}
    \epsfig{file=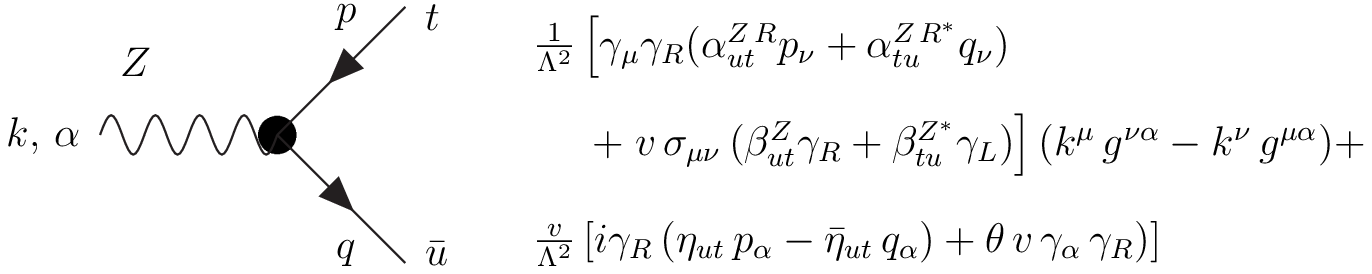,width=14 cm}
    \caption{Feynman rules for the anomalous vertex $Z \, t\, \bar{u} $.}
    \label{fig:feynZ}
  \end{center}
\end{figure}

\begin{figure}[htbp]
  \begin{center}
    \epsfig{file=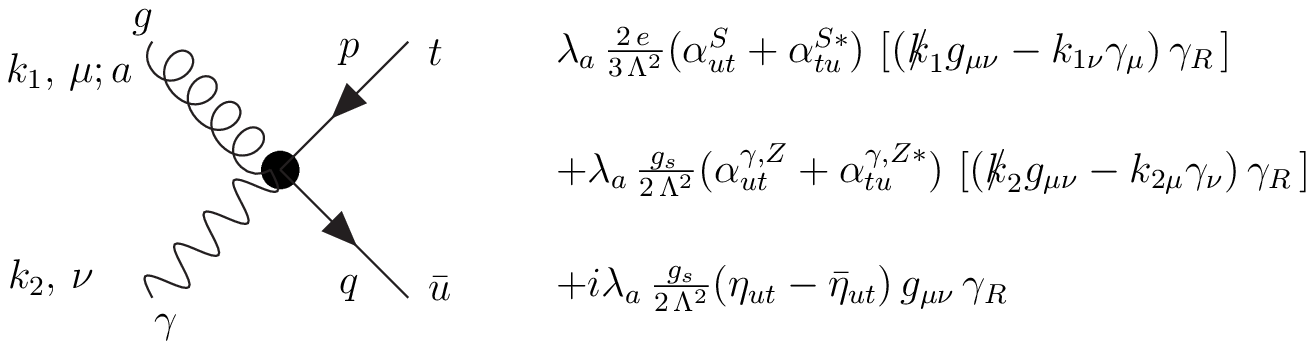,width=14 cm}
    \caption{Feynman rules for the anomalous quartic vertex $\gamma \, g\, t \bar{u} $.}
    \label{fig:feyn4gamma}
  \end{center}
\end{figure}

\begin{figure}[htbp]
  \begin{center}
    \epsfig{file=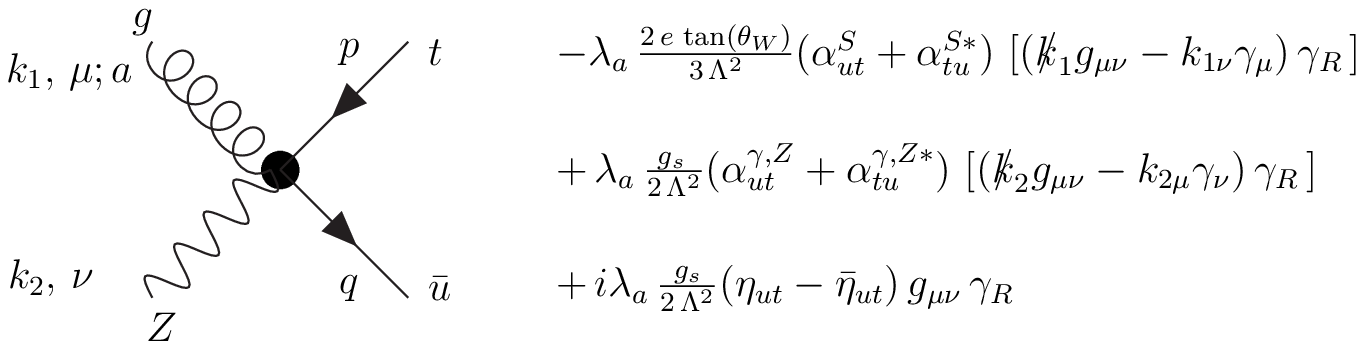,width=14 cm}
    \caption{Feynman rules for the anomalous quartic vertex $Z \, g\, t \bar{u} $.}
    \label{fig:feyn4Z}
  \end{center}
\end{figure}

The complete effective lagrangian can now be written as a function
of the operators defined in the previous section,
\begin{align}
{\cal L} &  = \,i \frac{\alpha^B_{it}}{\Lambda^2}\,\left(
\bar{u}^i_R \, \, \gamma_{\mu} D_{\nu} t_R \right) \, B^{\mu
\nu}\;+\;i \frac{\alpha^B_{ti}}{\Lambda^2}\,\left( \bar{t}_R \, \,
\gamma_{\mu}
D_{\nu} u^i_R \right) \, B^{\mu \nu} \nonumber \vspace{0.3cm} \\
 & \;+\; \frac{\beta^{W}_{it}}{\Lambda^2}\,\left(\bar{q}^i_L\, \,
\tau_{I} \, \sigma^{\mu\nu}\,t_R\right)\, \phi\,W^{I}_{\mu\nu}\; +
\;\frac{\beta^{W}_{ti}}{\Lambda^2}\,\left(\bar{t}_L\, \, \tau_{I} \,
\sigma^{\mu\nu}\,u^i_R\right)\, \tilde{\phi}\,W^{I}_{\mu\nu} \nonumber  \vspace{0.3cm} \\
 & \;+\; \frac{\beta^{B}_{it}}{\Lambda^2}\,\left(\bar{q}^i_L\,
\sigma^{\mu\nu}\,t_R\right)\, \tilde{\phi}\,B_{\mu\nu} \;+\;
\frac{\beta^{B}_{ti}}{\Lambda^2}\,\left(\bar{t}_L\,
\sigma^{\mu\nu}\,u^i_R\right)\, \phi\,B_{\mu\nu} \nonumber \vspace{0.3cm} \\
 & \;+\; \frac{\eta_{it}}{\Lambda^2}\,\left(\bar{q}^i_L\,
D^{\mu}\,t_R\right)\, D_{\mu} \tilde{\phi} \;+\;
\frac{\bar{\eta}_{it}}{\Lambda^2}\,\left( D^{\mu}
\bar{q}^i_L\, \,t_R\right)\, D_{\mu} \tilde{\phi} \nonumber \vspace{0.3cm} \\
 & \;+\; \theta_{it} \, (\phi^{\dagger} D_{\mu} \phi) \, (\bar{u^i_R} \gamma^{\mu}
 t_R) \; + \; \theta_{ti} \, (\phi^{\dagger} D_{\mu} \phi) \, (\bar{t_R} \gamma^{\mu}
 u^i_R)  \;+\; \mbox{h.c.} \;\;\; .
 \label{eq:lag}
\end{align}
This lagrangian describes new vertices of the form $\gamma
\,\bar{u}\, t$, $Z \,\bar{u}\, t$, $\bar{u} \,t \,\gamma \, g$ and
$\bar{u} \,t \, Z \, g$ (and many others) and their charge conjugate
vertices. For simplicity we redefine the $\eta$ and $\theta$
couplings as $ \eta \rightarrow (\sin (2 \theta_W)/e) \, \eta$ and $
\theta \rightarrow (\sin (2 \theta_W)/e) \, (\theta_{it}
\,-\,\theta^*_{ti})$. The Feynman rules for the FCNC triple vertices
are shown in figures~\eqref{fig:feyngamma}
and~\eqref{fig:feynZ}~\footnote{The Feynman rules for the
charge-conjugate vertices are obtained by simple complex
conjugation. The exception is the $\theta$ term, which due to our
definition of the $\theta$ coupling in eq.~\eqref{eq:op2}, will
become $-\theta^*$ for the vertex $Z \, u\, \bar{t}$.}.
Just like for the anomalous operators in the strong sector, the
gauge structure of the terms in eq.~\eqref{eq:lag} gives rise to new
quartic vertices. Most of the couplings which contribute to the
triple vertices of figs.~\eqref{fig:feyngamma},~\eqref{fig:feynZ}
also contribute to the quartic ones. The Feynman rules for the
quartic vertices we will need for this paper are shown in
figures~\eqref{fig:feyn4gamma} and~\eqref{fig:feyn4Z}. We see
that these quartic interactions receive contributions from both the
strong and electroweak effective operators. Their presence is
mandatory because of gauge invariance and they will be of great
importance to obtain several elegant results which we present in
section~\ref{sec:sig}.

For comparison, the FCNC lagrangian considered by the authors of
ref.~\cite{del Aguila:1999ec} consisted in
\begin{eqnarray}
\cal{L} &=\;
\displaystyle{\frac{g}{2\cos\theta_W}}\,\bar{t}\,\gamma_\mu\,(X_{tq}^L\,\gamma_L\,+\,
X_{tq}^R\,\gamma_R)\,q\,Z^\mu\;+\;\displaystyle{\frac{g}{2\cos\theta_W}}\,\bar{t}\,
(k_{tq}^{(1)}\,-\,i\,k_{tq}^{(2)}\gamma_5)\,\displaystyle{\frac{i\sigma_{\mu\nu}q^\nu}{m_t}}
\,q\,Z^\mu \nonumber \\
 & \;\;\;+\; e\,\bar{t}\,(\lambda_{tq}^{(1)}\,-\,i\,\lambda_{tq}^{(2)}\gamma_5)\,
 \displaystyle{\frac{i\sigma_{\mu\nu}q^\nu}{m_t}}\,q\,A^\mu\;+\;
g_S\,\bar{t}\,(\zeta_{tq}^{(1)}\,-\,i\,\zeta_{tq}^{(2)}\gamma_5)\,
 \displaystyle{\frac{i\sigma_{\mu\nu}q^\nu}{m_t}}\,T^a\,q\,G^{a \mu}\;+\;h.c.
 \label{eq:lagj}
 \end{eqnarray}
Notice that whereas we consider a generic scale $\Lambda$ for new
physics, these authors set $\Lambda\,=\,m_t$. Also, it is easy to
recognize several of our couplings in the lagrangian above; for
instance, we have
\begin{align}
\displaystyle{\frac{g}{2\cos\theta_W}}\,X_{tq}^R\,=\,
\displaystyle{\frac{v^2}{\Lambda^2}}\,\theta\;\;\; , &
\;\;\;\displaystyle{\frac{g}{4\cos\theta_W
m_t}}\,\left(k_{tq}^{(1)}\,-\,i\,k_{tq}^{(2)}\right)\,=\,
\displaystyle{\frac{v}{\Lambda^2}}\,\beta_{qt}^Z \;\;\; , \nonumber
\\ \displaystyle{\frac{\,e}{2 \, m_t} (\lambda_{tq}^{(1)}
\,-\,i \,\lambda_{tq}^{(2)})}
\,=\,\displaystyle{\frac{v}{\Lambda^2}} \, \beta_{qt}^\gamma \;\;\;
, & \;\;\; \displaystyle{\frac{g_S}{4 \, m_t} (\xi_{tq}^{(1)}-i \,
\xi_{tq}^{(2)})} \,=\, \displaystyle{\frac{v}{\Lambda^2}}
\,\beta_{qt}^S \;\;\; .
\end{align}
Notice that due to our choice of efective operators the couplings of
the form $\beta_{qt}$ and $\beta_{tq}$, and others, are treated as
independent - meaning, the lagrangian~\eqref{eq:lagj} does not
contain our couplings $\beta_{tq}$.  Also, couplings of the form
$\{\alpha\,,\,\eta\}$ are not present in~\eqref{eq:lagj}, and the
photon and $Z$ couplings therein presented are taken to be
completely independent, unlike what we considered in our work. Their
$X_{tq}^L$ coupling hasn't got an equivalent in our formulation. We
could obtain it through a $\theta$-like effective operator, namely,
\begin{equation}
(\phi^{\dagger} D_{\mu} \phi) \, (\bar{q^i_L} \gamma^{\mu} q^j_L)
\;\;\; ,
\end{equation}
where one of the quark doublets $q^i$, $q^j$ would contain the top
quark. It is easy to see, though, that this operator would have a
direct contribution to bottom quark physics, thus violating one of
our selection criteria for the anomalous top interactions. One
important remark: the authors of ref.~\cite{del Aguila:1999ec} do
not consider the quartic vertices of figs.~\eqref{fig:feyn4gamma}
and~\eqref{fig:feyn4Z}  in their calculations of cross sections for
$t\,+\,\gamma$ and $t\,+\,Z$ production. That's entirely correct,
since their analysis does not involve couplings like
$\{\alpha\,,\,\eta\}$, the only ones who contribute to those quartic
vertices.

\section{FCNC branching ratios of the top}
\label{sec:brs}

The top can have FCNC decays in the SM, but not at tree level. As
such, the branching ratios of these rare top decays are immensely
suppressed in the SM, but can be much larger in extensions of the
model. Essentially, the existence of new particles will give new
contributions to the top rare decays. The interesting thing is that
there can be differences of as much as thirteen orders of magnitude
between the SM branching ratios and those in some models, as may be
seen in table~\ref{tab:br}.
\begin{table}[htb]
\vspace*{0.2cm}
\begin{center}
\begin{small}
\begin{tabular}{lccccc}
\hline  \hline
 Process  & SM & QS & 2HDM & MSSM & $R \!\!\!\!\!\!  \not \quad$ SUSY
   \\
\hline  \\[-0.5cm]
$t \to u Z$ & $8 \times 10^{-17}$ & $1.1 \times 10^{-4}$
  & $-$
  & $2 \times 10^{-6}$ & $3 \times 10^{-5}$ \\
$t \to u \gamma$ & $3.7 \times 10^{-16}$ & $7.5 \times 10^{-9}$
 & $-$
 & $2 \times  10^{-6}$ & $1 \times 10^{-6}$ \\
$t \to u g$ & $3.7 \times 10^{-14}$ & $1.5 \times 10^{-7}$
  & $-$
  & $8 \times 10^{-5}$ & $2 \times 10^{-4}$ \\
[0.2cm]
$t \to c Z$ & $1 \times 10^{-14}$ & $1.1 \times 10^{-4}$
  & $\sim 10^{-7}$
  & $2 \times 10^{-6}$ & $3 \times 10^{-5}$ \\
$t \to c \gamma$ & $4.6 \times 10^{-14}$ & $7.5 \times 10^{-9}$
 & $\sim 10^{-6}$
 & $2 \times 10^{-6}$ & $1 \times 10^{-6}$ \\
$t \to c g$ & $4.6 \times 10^{-12}$ & $1.5 \times 10^{-7}$
  & $\sim 10^{-4}$
  & $8 \times 10^{-5}$ & $2 \times 10^{-4}$ \\
\hline \hline
\end{tabular}
\end{small}
\end{center}
\caption{Branching ratios for FCNC decays of the top quark in the SM
and several possible extensions: the quark-singlet model (QS), the
two-higgs doublet model (2HDM), the minimal supersymmetric model
(MSSM) and SUSY with R-parity violation. See
ref.~\cite{AguilarSaavedra:2004wm, calc} for details.}
\label{tab:br}
\end{table}
The effective operator formalism allows us to describe, in a
model-independent manner, the possible rare decays of the top. In
ref.~\cite{Ferr1} we computed the branching ratios for the FCNC top
decays $t\,\rightarrow\,q\,g$, due to the strong sector anomalous
operators therein introduced. The decay width for
$t\,\rightarrow\,u\,g$ is given by
\begin{align}
\Gamma (t \rightarrow u g) &=\;  \frac{m^3_t}{12
\pi\Lambda^4}\,\Bigg\{ m^2_t \,\left|\alpha_{tu}^S  +
(\alpha^S_{ut})^* \right|^2 \,+\, 16 \,v^2\, \left(\left|
\beta_{tu}^S \right|^2 + \left| \beta_{ut}^S \right|^2 \right)
\;\;\; +
\vspace{0.3cm} \nonumber \\
 & \hspace{2.2cm}\, 8\, v\, m_t\,\mbox{Im}\left[ (\alpha_{ut}^S  + (\alpha^S_{tu})^*)
\, \beta_{tu}^S \right] \Bigg\} \label{eq:widS}\;\;\; ,
\end{align}
with an analogous expression for $\Gamma (t \rightarrow c g)$, with
different couplings. The electroweak sector operators we discussed
in the previous section contribute to new FCNC decays, namely,
$t\,\rightarrow \,u\,\gamma$ (and $t\,\rightarrow \,c\,\gamma$, with
{\em a priori} different couplings), for which we obtain a width
given by the following expression:
\begin{align}
\Gamma (t \rightarrow u \gamma) &=\;  \frac{m^3_t}{64
\pi\Lambda^4}\,\Bigg\{ m^2_t \,\left|\alpha_{tu}^{\gamma}  +
(\alpha^{\gamma}_{ut})^* \right|^2 \,+\, 16 \,v^2\, \left(\left|
\beta_{tu}^{\gamma} \right|^2 + \left| \beta_{ut}^{\gamma} \right|^2
\right) \;\;\; +
\vspace{0.3cm} \nonumber \\
 & \hspace{2.2cm}\, 8\, v\, m_t\,\mbox{Im}\left[ (\alpha_{ut}^{\gamma}  + (\alpha^{\gamma}_{tu})^*)
\, \beta_{tu}^{\gamma} \right] \Bigg\} \label{eq:widW} \;\;\; .
\end{align}
Notice how similar this result is to eq.~\eqref{eq:widS}. We will
also have contributions from these operators to $t\,\rightarrow
\,u\,Z$ ($t\,\rightarrow \,c\,Z$), from which we obtain a width
given by
\begin{eqnarray}
 \Gamma(t\,\rightarrow \,u\,Z) & = & \frac{{\left( m_t^2 - m_Z^2 \right) }^2}{32\,m_t^3\,\pi
\,\Lambda^4}
\left[ K_1 \, \left| \alpha^Z_{ut} \right|^2 + K_2 \, \left|
\alpha^Z_{tu} \right|^2 + K_3 \, ( \left| \beta^Z_{ut} \right|^2 +
\left| \beta^Z_{tu} \right|^2)+ K_4 \, ( \left| \eta_{ut} \right|^2
+ \left| \bar{\eta}_{ut} \right|^2) \right.
\nonumber \\[0.25cm]
&&  \qquad + \, K_5 \, \left| \theta \right|^2 + K_6 \, Re \left[
\alpha^Z_{ut} \, \alpha^Z_{tu} \right] + K_7 \, Im \left[
\alpha^Z_{ut} \, \beta^Z_{tu} \right]
\nonumber \\[0.25cm]
&&  \qquad + \, K_8 \, Im \left[ \alpha^{Z^*}_{tu} \, \beta^Z_{tu}
\right] + K_9 \, Re \left[ \alpha^Z_{ut} \theta^* \right]+ K_{10} \,
Re \left[ \alpha^Z_{tu} \theta \right]
\nonumber \\[0.25cm]
&& \qquad   \left. + \, K_{11} \, Re \left[ \beta^Z_{ut}
(\eta_{ut}-\bar{\eta}_{ut})^* \right] + K_{12} \, Im \left[
\beta^Z_{tu} \, \theta \right] + K_{13} \, Re \left[ \eta_{ut}
\bar{\eta}_{ut}^* \right] \right]\;\;\; ,
\end{eqnarray}
where the coefficients $K_i$ are given by
\begin{eqnarray}
K_1 & = & \frac{1}{2} \, (m_t^4 + 4\,m_t^2\,m_Z^2 + m_Z^4) \qquad
K_2 \, = \, \frac{1}{2} \,  (m_t^2 - m_Z^2)^2 \qquad K_3 \, = \,
4\,( 2\,m_t^2 + m_Z^2) \,v^2
 \nonumber  \\
K_4 & = & \frac{v^2}{4\,m_Z^2}(m_t^2 - m_Z^2)^2 \qquad K_5 \, = \,
\frac{v^4}{m_Z^2} ( m_t^2 + 2\,m_Z^2 ) \qquad K_6 \, = \, ( m_t^2 -
m_Z^2 ) \,( m_t^2 + m_Z^2)
\nonumber \\
K_7 & = & 4\,m_t\, ( m_t^2 + 2\,m_Z^2 ) \,v \qquad K_8 \, = \,
4\,m_t\, ( m_t^2 - m_Z^2 ) \,v \qquad K_9 \, = \, -2\, ( 2\,m_t^2 +
m_Z^2 ) \,v^2
\nonumber \\
K_{10} & = & -2\, ( m_t^2 - m_Z^2) \,v^2 \qquad K_{11} \, = - K_{10}
\qquad K_{12} \, = \, -12 \,m_t\,v^3 \qquad K_{13} \, = \, \frac{-
v^2 }{m_Z^2} \, K_2 \;\;\;\ .
\end{eqnarray}
\begin{table}[t]
\begin{center}
  \begin{tabular}{ | l | c | c | c |}
    \hline
     & LEP & HERA & Tevatron  \\ \hline \hline
    $Br(t \rightarrow q \, Z)$      & $ < \, 7.8 \% \,$~\cite{LEP2Zgamma} & $  < \, 49\% \,$~\cite{Zeus}
    & $ < \, 10.6 \% \,^d$~\cite{tZqCDF}  \\ \hline
    $Br(t \rightarrow q \, \gamma)$ & $ < \, 2.4 \% \,$~\cite{LEP2Zgamma} & $ < \, 0.75 \% \,$~\cite{Zeus}
    & $ < \, 3.2 \% \,^d$~\cite{gammaCDF}  \\ \hline
    $Br(t \rightarrow q \, g)$      & $ < \, 17 \% \,$~\cite{YR1}  & $ < \, 13 \% \,$~\cite{Ashimova:2006zc,Zeus}
    & $ < \, O (0.1 - 1 \%) \,$~\cite{gluonTevatron} \\
    \hline
  \end{tabular}
\end{center}
\caption{Current experimental bounds on FCNC branching ratios. The
upperscript ``d" refers to bounds obtained from direct measurements,
as is explained in the text.} \label{tab:limits}
\end{table}

There are several experimental bounds for FCNC processes. As we
mentioned earlier, indirect bounds \cite{Fox:2007in,EPM} originate
from electroweak precision physics and from B and K physics. The
strongest bounds so far are the ones in \cite{Fox:2007in} where
invariance under $SU(2)_L$ is required for the set of operators
chosen. This way top and bottom physics are related and B physics
can be used to set limits on operators that involve top and bottom
quarks through gauge invariance. Regarding $Br (t \,\rightarrow\,
q\, Z)$ and $Br (t\,\rightarrow \,q\, \gamma)$, the only direct
bounds available to date are the ones from the Tevatron (CDF). The
CDF collaboration has searched its data for signatures of $t
\,\rightarrow \,q \, \gamma$ and $t\, \rightarrow \,q \, Z$ (where
$q\,=\,u,c$). Both analyses use $p\bar{p}\, \rightarrow \,
t\,\bar{t} $ data and assume that one of the tops decays according
to the SM into $W\,b$. The results are presented in Table
\ref{tab:limits}. As data is still being collected, we expect that
these bounds will improve in the near future. The bounds on the
branching ratios from LEP and ZEUS are bounds on the cross section
that were then translated into bounds on the branching ratios
through the anomalous couplings. The LEP bounds use the same
anomalous coupling for the $u$ and $c$ quarks and the ZEUS bound is
only for the process involving a $u$ quark. The bounds on $Br (t
\rightarrow \,q\, g)$ are all from cross sections translated into
branching ratios. Usually only one operator is considered, the
chromomagnetic one, which makes the translation straightforward. The
same searches are being prepared for the LHC. A detailed discussion
with all present bounds on FCNC and the predictions for the LHC can
be found in \cite{toni, fla, CMS}. With a luminosity of 100
$fb^{-1}$ and in the absence of signal, the 95\% confidence level
bounds on the branching ratios give us $Br(t \,\rightarrow\, q\,
Z)\,\sim\,10^{-5}$, $Br(t \,\rightarrow\, q\,
\gamma)\,\sim\,10^{-5}$ and $Br(t \,\rightarrow\, q\,
g)\,\sim\,10^{-4}$.

Let us now recall that the anomalous couplings that describe the
FCNC decays $t \,\rightarrow\, q\, Z$ and $t \,\rightarrow\, q\,
\gamma$ are not entirely independent - according to
eqs.~\eqref{eq:alf} and~\eqref{eq:bet} the couplings
$\{\alpha^\gamma\,,\,\alpha^Z\}$ and $\{\beta^\gamma\,,\,\beta^Z\}$
are related to one another. This will imply a correlation of sorts
between the branching ratios for these two decays. Then, gauge
invariance imposes that one can consider anomalous FCNC interactions
that affect only the decay $t \,\rightarrow\, q\, Z$, but any
anomalous interactions which affect $t \,\rightarrow\, q\, \gamma$
will necessarily have an impact on $t \,\rightarrow\, q\, Z$. In
particular, if one considers any sort of theory for which $Br(t
\,\rightarrow\, q\, \gamma)\,\neq\,0$, then one will forcibly have
$Br(t \,\rightarrow\, q\, Z)\,\neq\,0$. The reverse of this
statement is not necessarily true, since more anomalous couplings
contribute to the $Z$ interactions than do the $\gamma$ ones.

\begin{figure}[htbp]
  \begin{center}
    \epsfig{file=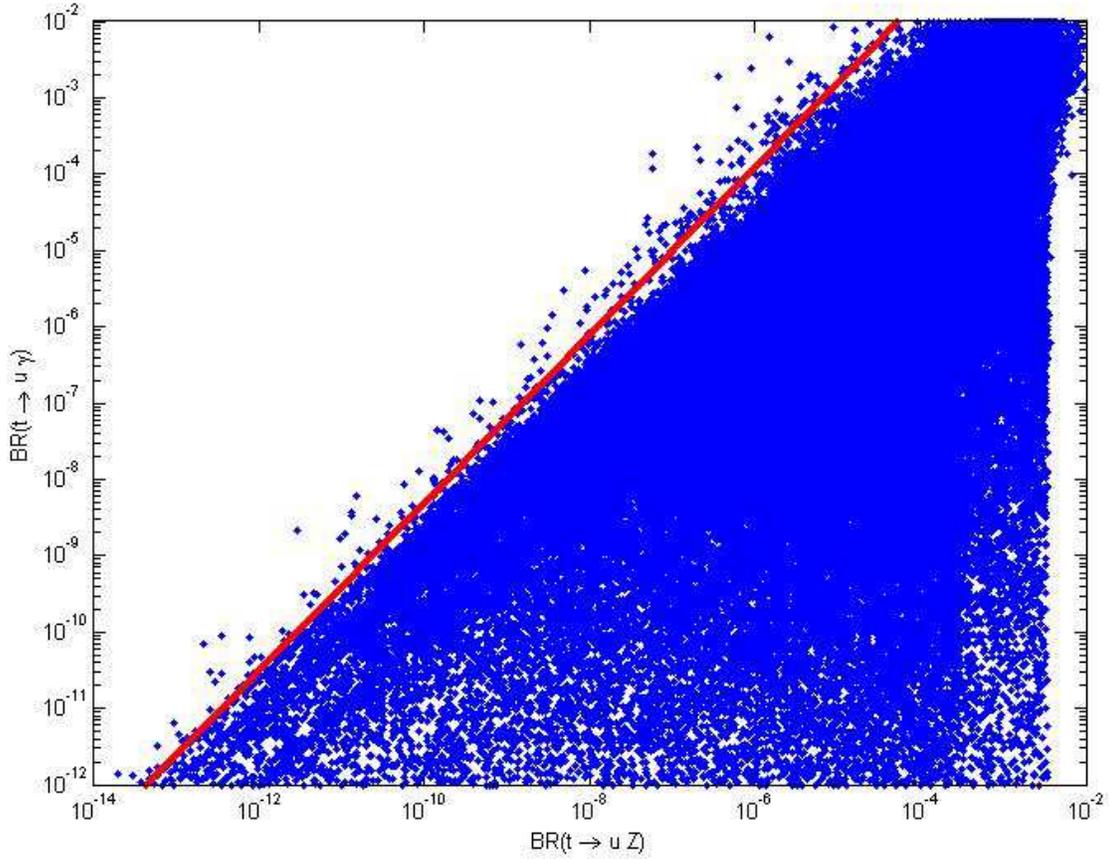,width=17 cm}
    \caption{FCNC branching ratios for the decays $t \,\rightarrow\, u\, Z$ vs.
    $t \,\rightarrow\, u\, \gamma$. The straight line corresponds to $500\times
    [Br(t \,\rightarrow\,u\,Z)]^{1.1}$.}
    \label{fig:brs}
  \end{center}
\end{figure}

If the couplings contributing to one of these branching ratios were
completely unrelated to those contributing to the other, then the
two branching ratios would be completely independent of one another.
As we see in figure~\eqref{fig:brs} that is not the case. To obtain
this plot we considered that the total width of the top quark was
equal to $1.42$ GeV (a value which includes QCD corrections, and
taking $V_{tb}\,\simeq\,1$~\cite{YR1,qcdc}), set $\Lambda\,=\,1$
TeV~\footnote{If one wishes to consider a different scale for new
physics, one will simply have to rescale the values of the anomalous
couplings.} and generated random complex values of all the anomalous
couplings, with magnitudes in the range between $10^{-10}$ and $1$.
We rejected those combinations of parameters which resulted in
branching ratios for $t \,\rightarrow\, u\, Z$ and $t
\,\rightarrow\, u\, \gamma$ larger than $10^{-2}$~\footnote{With all
precision one should then add the corresponding FCNC widths to the
top total width quoted above. However, the error we commit with this
approximation is always smaller than 2\%, and then only for the
larger values of the branching ratios considered.}. Regarding the
$\{\alpha\,,\,\beta\}$ couplings, we first generated random values
for $\{\alpha_{ij}^B\,,\,\beta_{ij}^B\,,\,\beta_{ij}^W\}$ and then,
through eqs.~\eqref{eq:alf} and~\eqref{eq:bet} obtained
$\{\alpha^\gamma\,,\,\alpha^Z\}$ and $\{\beta^\gamma\,,\,\beta^Z\}$.

With very little exceptions, we can even quote a rough bound on the
branching ratios by observing the straight line drawn by us in the
plot - namely, that it is nearly impossible to have $Br(t
\,\rightarrow\, u\, \gamma)\,>\,500\,Br(t \,\rightarrow\,
u\,Z)^{1.1}$. Again, if gauge invariance did not impose the
conditions between $\gamma$ and $Z$ couplings expressed in
eqs.~\eqref{eq:alf} and~\eqref{eq:bet}, what we would obtain in
fig.~\eqref{fig:brs} would be a uniformly filled plot - for a given
value of $Br(t \,\rightarrow\, u\, Z)$ one could have any value of
$Br(t \,\rightarrow\, u\, \gamma)$. If we take the point of view
that any theory beyond the SM will manifest itself at the TeV scale
through the effective operators of ref.~\cite{buch} then this
relationship between these two FCNC branching ratios of the top is a
model-independent prediction. Finally, had we considered a more
limited set of anomalous couplings - for instance, only $\alpha$ or
$\beta$ type couplings - the plot in fig.~\eqref{fig:brs} would be
considerably simpler. Due to the relationship between those
couplings, the plot would reduce to a band of values, not a wedge as
that shown. Identical results were obtained for the FCNC decays $t
\,\rightarrow\, c\, Z$ and $t \,\rightarrow\, c\, \gamma$.

\section{Strong vs. Electroweak FCNC contributions for cross sections of
associated single top production} \noindent \label{sec:sig}

\begin{figure}[htbp]
  \begin{center}
    \epsfig{file=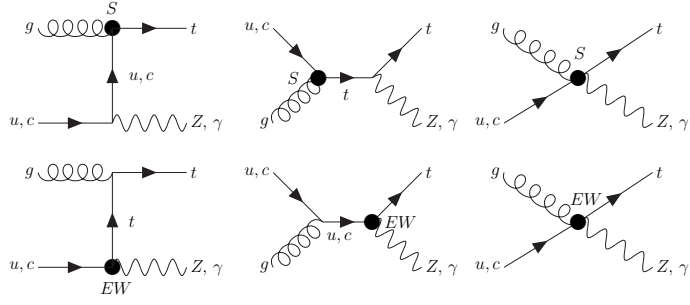,width=10 cm}
    \caption{Feynman diagrams for $t \, Z$ and $t \, \gamma$ production
    with both strong and electroweak FCNC vertices.}
    \label{fig:gqtZgamma}
  \end{center}
\end{figure}

The anomalous operators considered in this paper contribute, not
only to FCNC decays of the top, but also to processes of single top
production. Namely to the associated production of a top quark
alongside a photon or a $Z$ boson, processes described by the
Feynman diagrams shown in fig.~\eqref{fig:gqtZgamma}. The FCNC
vertices are represented by a solid dot, with the letter ``S"
standing for a strong FCNC anomalous interaction and a ``EW" for the
electroweak one. Notice the four-legged diagrams, imposed by gauge
invariance. The strong-FCNC channels had already been considered in
ref.~\cite{Ferr3}. Our aim in this section is to investigate what is
the combined influence of the strong and electroweak anomalous
contributions to these processes.

\subsection{Cross section for $q \, g \rightarrow t \, \gamma$}

\begin{figure}[htbp]
  \begin{center}
    \epsfig{file=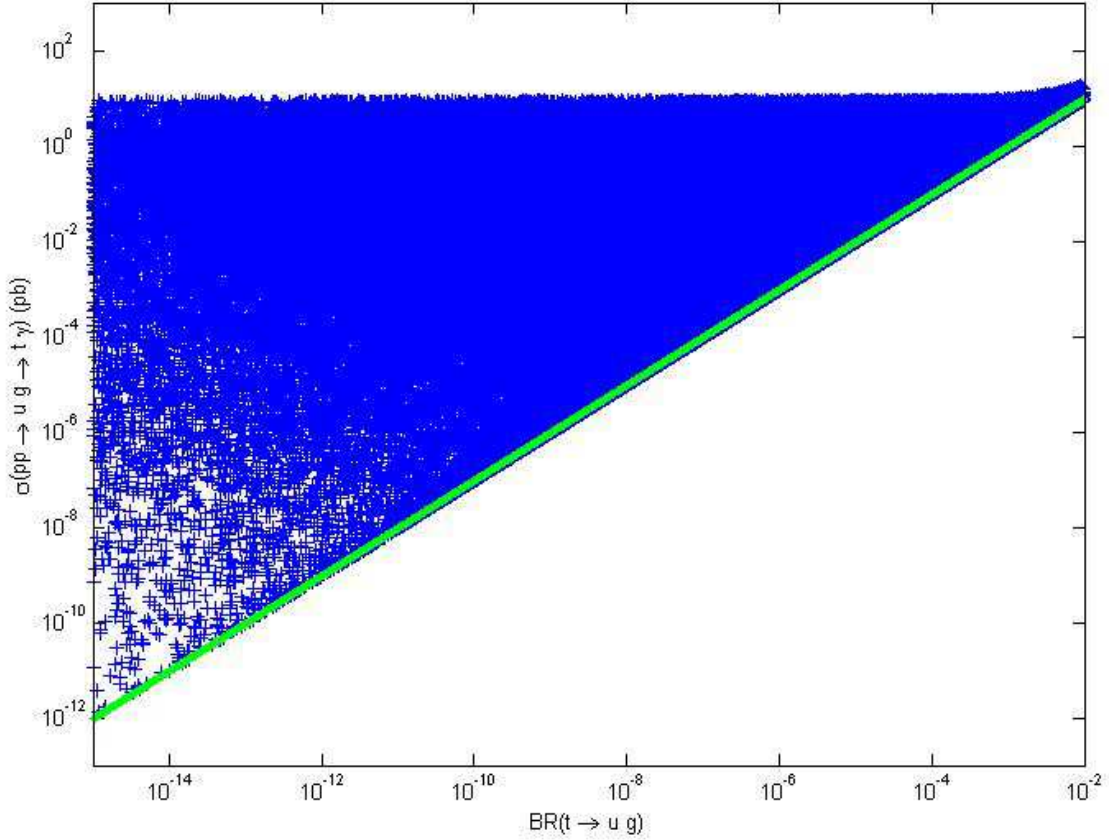,width=17 cm}
    \caption{Total (black crosses) and strong (grey) cross
    sections for the process $p\,p\rightarrow\,u\,g\rightarrow t\,\gamma$ versus the
    FCNC branching ratio for the decay $t\,\rightarrow\,u\,g$.}
    \label{fig:sigtga}
  \end{center}
\end{figure}

The total cross section for the associated FCNC production of a
single top quark and a photon including all the anomalous
interactions considered in section~\ref{sec:eff} is given by
\begin{eqnarray}
\frac{d \,\sigma_{q \, g \rightarrow t \, \gamma}}{dt} & = &
\frac{e^2}{18\,m_t^3\,s^2}\, F_{\gamma}(t,s) \,
\Gamma(t\,\rightarrow\,q\,g) \,+\, \frac{g_S^2}{6\,m_t^3\,s^2}\,
F_{\gamma}(s,t) \, \Gamma(t\,\rightarrow\,q\,\gamma) \;+\; \frac{e
\, g_S\,H_{\gamma}(t,s)}
  {96 \, \pi \, s^2 \, \Lambda^4} \,\times \nonumber \\[0.3cm]
&&  \left\{ Re \left[ \left( \alpha^S_{it} + (\alpha^S_{ti})^*
\right) \left( \alpha^{\gamma}_{it} + (\alpha^{\gamma}_{ti})^*
\right) \right] + \frac{4v}{m_t} \, Im \left[
((\alpha^\gamma_{it})^* + \alpha^\gamma_{ti}) \, \beta^S_{ti} +
(\alpha^S_{it} + ( \alpha^S_{ti})^*) \,
\beta^\gamma_{ti} \right] \right. \nonumber \\[0.3cm]
&& \qquad + \left. \frac{16v^2}{m_t^2} \, Re \left[
\beta^\gamma_{it} (\beta^S_{it})^*  + \beta^\gamma_{ti}
(\beta^S_{ti})^* \right] \right\} \label{eq:tgam}
\end{eqnarray}
where we have defined the functions
\begin{eqnarray}
F_{\gamma}(t,s) & = & \frac{{m_t}^8 + 2\,s^2\,t\,\left( s + t
\right) -      {m_t}^6\,\left( s + 2\,t \right)  +
      {m_t}^4\,\left( s^2 + 4\,s\,t + t^2 \right)  -
      {m_t}^2\,s\,\left( s^2 + 6\,s\,t + 3\,t^2 \right)}{\left( {m_t}^2 - s \right)^2\,t}
\nonumber \\
H_{\gamma}(t,s)  & = &  -\,\frac{2\, m_t^2}{3\,
    \left( m_t^2 - s \right) \,\left( m_t^2 - t \right) } \left(
3\,m_t^6 -
      4\,m_t^4\,\left( s + t \right)  - s\,t\,\left( s + t \right)  +
      m_t^2\,\left( s^2 + 3\,s\,t + t^2 \right)  \right) \;\;\; .
\end{eqnarray}
We used the couplings generated in the previous section for which we
computed the branching ratios presented in fig.~\eqref{fig:brs}. We
also generated random complex values for the strong couplings
$\{\alpha_{ij}^S\,,\,\beta_{ij}^S\}$, once again requiring that
$Br(t\,\rightarrow\,u\,g)\,<\,10^{-2}$. To obtain the cross section
for the process $p\,p\,\rightarrow\,u\,g\,\rightarrow\,t\,\gamma$ at
the LHC we integrated the partonic cross section in
eq.~\eqref{eq:tgam} with the CTEQ6M partonic distribution
functions~\cite{cteq6}, with a factorization scale $\mu_F$ set equal
to $m_t$. We also imposed a cut of 10 GeV on the $p_T$ of the final
state partons. In figure~\eqref{fig:sigtga} we plot the value of the
cross section for this process against the branching ratio of the
FCNC decay of the top to a gluon. We show both the ``strong" cross
section (in grey, corresponding to all couplings but the strong ones
set to zero) and the total cross section (in black crosses,
including the effects of the strong couplings, the electroweak ones
and their interference). The most immediate conclusion one can draw
from
fig.~\eqref{fig:sigtga} is that the interference between the strong
and weak FCNC interactions is by and large constructive. In fact,
the vast majority of the points in fig.~\eqref{fig:sigtga} which
correspond to the total cross section lie above the line
representing the contributions from the strong FCNC processes alone.
For a small subset of points we may have
$\sigma^{Total}(pp\rightarrow u g\rightarrow
t\gamma)\,<\,\sigma^{S}(pp\rightarrow u g\rightarrow t\gamma)$, but
in those cases the difference between both quantities is never
superior to 1\%. Then, within an error of 1\%, the strong cross
section $\sigma^{S}(pp\rightarrow u g\rightarrow t\gamma)$
(calculated in ref.~\cite{Ferr3}) is effectively a lower bound on
the total cross section for this process.

Another interesting observation from fig.~\eqref{fig:sigtga}: any
bound on $Br(t\,\rightarrow\,u\,g)$ (such as those which are
expected to come from the LHC results) immediately implies a bound
on $\sigma(p\,p\rightarrow u g\rightarrow t\,\gamma)$ - and
vice-versa. However, a hypothetical direct determination of
$Br(t\,\rightarrow\,u\,g)$ would not determine the cross section, it
would only provide us with a lower bound on $\sigma(p\,p\rightarrow
u g\rightarrow t\,\gamma)$. Inversely, the discovery of the FCNC
process $p\,p \rightarrow u g\rightarrow t\,\gamma$ and obtention of
a value for $\sigma(p\,p\rightarrow u g\rightarrow t\,\gamma)$ would
set an upper bound on $Br(t\,\rightarrow\,u\,g)$, not fix its value.

\begin{figure}[htbp]
  \begin{center}
    \epsfig{file=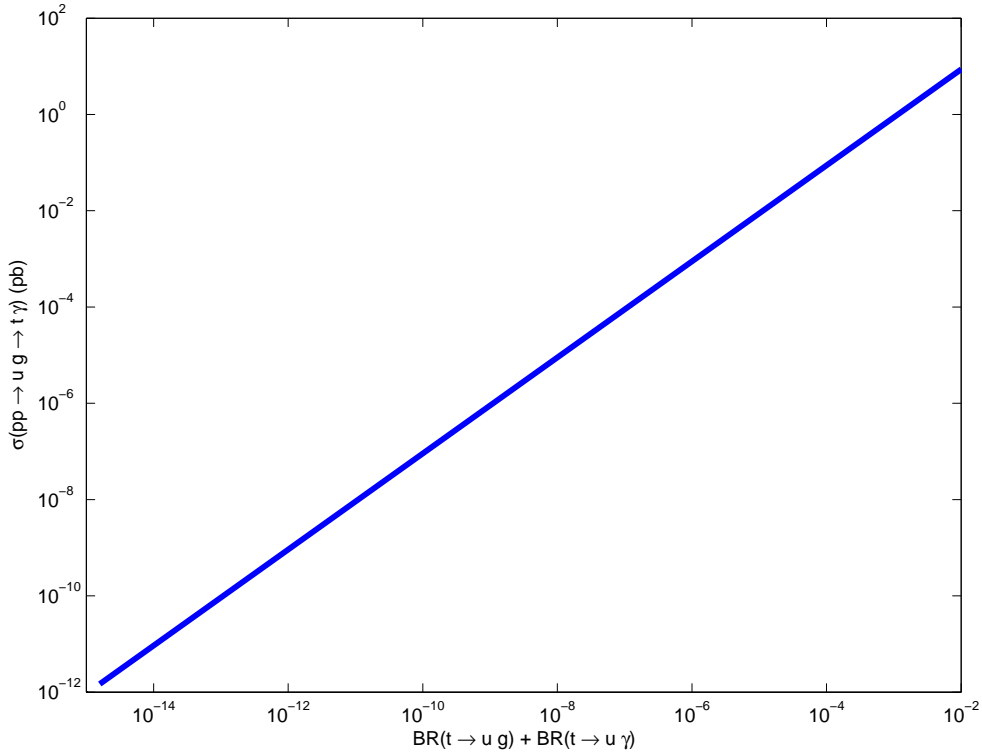,width=15 cm}
    \caption{Total (electroweak and strong contributions) cross
    section for the process $p\,p\rightarrow\,u\,g\rightarrow t\,\gamma$ versus
    the sum of the FCNC branching ratios for the decays
    $t\,\rightarrow\,u\,\gamma$ and $t\,\rightarrow\,u\,g$.}
    \label{fig:sigbrs}
  \end{center}
\end{figure}

Had we plotted the electroweak cross section (the term proportional
to $\Gamma(t\,\rightarrow\,q\,\gamma)$ in eq.~\eqref{eq:tgam}) and
the total one versus $Br(t\,\rightarrow\,u\,\gamma)$, we would have
found a very similar picture to that of fig.~\eqref{fig:sigtga}: a
straight line for the electroweak cross section and a wedge of
values lying mostly above it. Again, to within 1\% of the value of
the cross sections, the electroweak cross section
$\sigma^{EW}(pp\rightarrow u g\rightarrow t\gamma)$ is a lower bound
for the complete cross section. And as before, knowing the value of
$Br(t\,\rightarrow\,u\,\gamma)$ sets only a lower bound on
$\sigma(p\,p\rightarrow u g\rightarrow t\,\gamma)$, and determining
a value for the cross section establishes an upper bound on the
branching ratio. We thus observe a great similarity in the behaviour
of the total cross sections with both FCNC branching ratios. In
fact, this is shown in quite an impressive manner in
fig.~\eqref{fig:sigbrs}, where we
plot the total cross section against the {\em sum} of the FCNC
branching ratios. The ``line" shown in this figure is actually a
very thin band, but this plot shows that, to good approximation, we
should expect a direct proportionality between the cross section for
the process $p\,p \rightarrow u\,g \rightarrow t\,\gamma$ and the
quantity
$Br(t\,\rightarrow\,u\,\gamma)\,+\,Br(t\,\rightarrow\,u\,g)$. In
fact we can even extract the proportionality constant from the plot
above, and obtain
\begin{equation}
\sigma(p\,p \rightarrow u\,g \rightarrow t\,\gamma)\;\simeq\;900\,
\left[Br(t\,\rightarrow\,u\,\gamma)\,+\,Br(t\,\rightarrow\,u\,g)\right]
\;\;\mbox{pb}\;\;\;,
\end{equation}
with a maximal deviation of about 9\%. Thus a measurement of this
cross section would determine the {\em sum} of the FCNC branching
ratios, but not each of them separately. Analogous results are
obtained for the processes involving the $c$ quark, the only
differences stemming from the parton density functions associated
with that particle. We obtain
\begin{equation}
\sigma(p\,p \rightarrow c\,g \rightarrow t\,\gamma)\;\simeq\;95\,
\left[Br(t\,\rightarrow\,c\,\gamma)\,+\,Br(t\,\rightarrow\,c\,g)\right]
\;\;\mbox{pb}\;\;\;,
\end{equation}
but the values of the cross section can now deviate as much as 19\%
from this formula. Notice that typical values of the cross section
for production of $t\,+\,Z$ via FCNC through a $c$ quark are roughly
ten times smaller than those of processes that go through a $u$
quark, which is of course due to the much smaller charm content of
the proton.

\begin{figure}[htbp]
  \begin{center}
    \epsfig{file=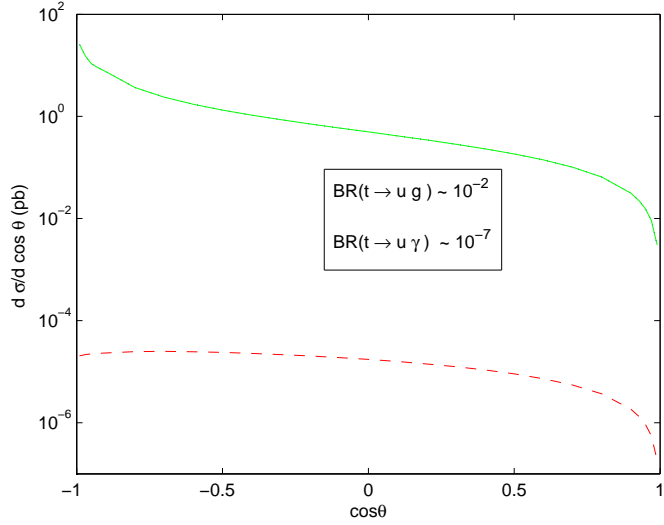,width=10 cm}
    \caption{Differential cross section $p\,p\rightarrow\,u\,g\rightarrow t\,\gamma$ versus
    $\cos\theta$, for a typical choice of parameters with a branching ratio
    for $t\,\rightarrow\,u\,g$ much larger than $Br(t\,\rightarrow\,u\,\gamma)$. The strong
    contribution practically coincides with the total cross section (full line). The electroweak
    contribution is represented by the dashed line.}
    \label{fig:diff1}
  \end{center}
\end{figure}

Is there a way, then, to ascertain whether the main contribution to
$\sigma(p\,p\rightarrow u g\rightarrow t\,\gamma)$ stems from
anomalous strong interactions, or from weak ones? Indeed there is,
by analysing the differential cross section for this process. In
fig.~\eqref{fig:diff1} we plot $d\sigma/d\cos\theta$ versus
$\cos\theta$, $\theta$ being the angle between the momentum of the
photon (or top) and the beam line. We show the strong and
electroweak contributions to this cross section, as well as its
total result. We chose a typical set of values for the anomalous
couplings producing a branching ratio for the FCNC decay
$t\,\rightarrow\,u\,g$ clearly superior to that of the decay
$t\,\rightarrow\,u\,\gamma$. As we see, the angular distribution of
the electroweak and strong cross sections is quite different. Since
the strong anomalous interactions are dominating over the
electroweak ones the total cross section mimics very closely the
strong one.

\begin{figure}[htbp]
  \begin{center}
    \epsfig{file=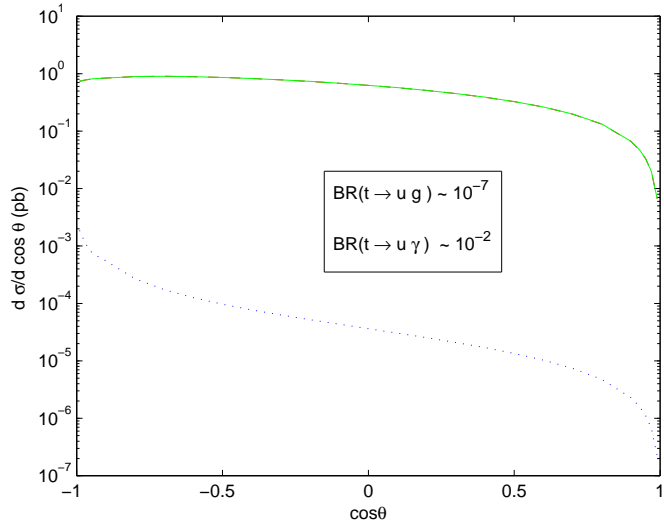,width=10 cm}
    \caption{Differential cross section $p\,p\rightarrow\,u\,g\rightarrow t\,\gamma$ versus
    $\cos\theta$, for a typical choice of parameters with a branching ratio
    for $t\,\rightarrow\,u\,g$ much smaller than $Br(t\,\rightarrow\,u\,\gamma)$. The
    electroweak contribution practically coincides with the total cross section (full line). The
    strong contribution is represented by the dotted line.}
    \label{fig:diff2}
  \end{center}
\end{figure}
\begin{figure}[htbp]
  \begin{center}
    \epsfig{file=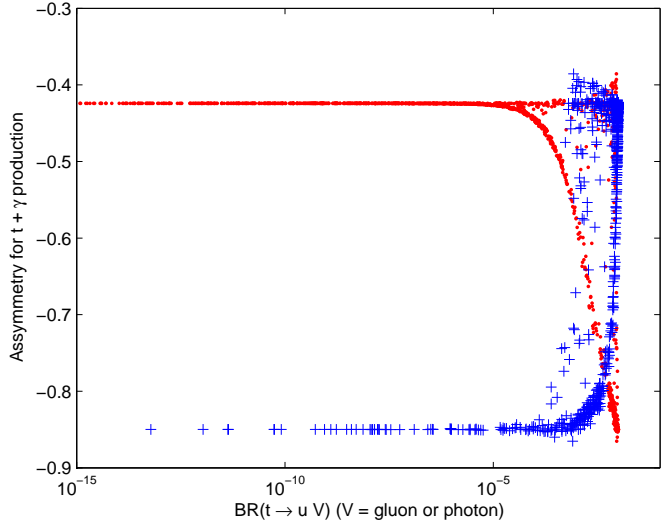,width=10 cm}
    \caption{The angular asymmetry coefficient defined in eq.~\eqref{eq:assg}
    as a function of the branching ratios $Br(t\,\rightarrow\,u\,\gamma)$ (crosses)
    and $Br(t\,\rightarrow\,u\,g)$ (dots).}
    \label{fig:assg}
  \end{center}
\end{figure}

In fig.~\eqref{fig:diff2} we show the inverse situation: a typical
set of values was chosen which gives us
$Br(t\,\rightarrow\,u\,\gamma)\,\sim\,10^{-2}$ and
$Br(t\,\rightarrow\,u\,g)\,\sim\,10^{-7}$, meaning a situation for
which the anomalous electroweak interactions are clearly dominant
over the strong ones. We see from the angular distribution of the
total cross section shown in fig.~\eqref{fig:diff2} that it now
greatly resembles its electroweak component. Judging from
figs.~\eqref{fig:diff1} and~\eqref{fig:diff2}, the telltale sign of
dominance of strong FCNC interactions is a pronounced variation with
$\cos\theta$ in the cross section, whereas a dominance of
electroweak FCNC effects will produce a relatively ``flat" cross
section. The Feynman diagrams of fig.~\eqref{fig:gqtZgamma} help to
explain this difference in dependence with $\cos\theta$: the strong
cross section has a significant contribution from the $t$-channel
(since the $s$-channel diagram is suppressed by the top mass),
whereas the inverse happens for the electroweak cross section.
However, it should be pointed out that the four-legged diagrams
contributing to both cross sections will upset a clear $s$-or-$t$
channel dominance. Notice also that if FCNC produce branching ratios
of similar size in both sectors the difference in behaviour shown in
these plots will not be seen. In fact, we may get a better feel for
the different angular behavior of the strong and electroweak FCNC
interactions if we define an asymmetry coefficient for this cross
section,
\begin{equation}
A_{t + \gamma} \;=\;\frac{\sigma_{t + \gamma}(\cos\theta > 0)
\,-\,\sigma_{t + \gamma}(\cos\theta < 0)}{\sigma_{t +
\gamma}(\cos\theta > 0) \,+\,\sigma_{t + \gamma}(\cos\theta <
0)}\;\;\; . \label{eq:assg}
\end{equation}
To exemplify the relevance of this quantity, we generated a special
sample of anomalous couplings: random values of all strong and
electroweak couplings such that
$Br(t\,\rightarrow\,u\,\gamma)\,+\,Br(t\,\rightarrow\,u\,g) \sim
10^{-2}$. This will include the cases where one of the branching
ratios dominates over the other, and also the case where both of
them have similar magnitudes. We show the results in
fig.~\eqref{fig:assg}, plotting the value of $A_{t + \gamma}$ in
terms of the two branching ratios whose sum is fixed to $10^{-2}$.
Looking at the far left of the plot we see that when the electroweak
FCNC interactions dominate over the strong ones $A_{t + \gamma}$
tends to a value of approximately $-0.85$, and in the reverse
situation we have $A_{t + \gamma}\,\sim\,-0.42$. However, when both
branching ratios have similar sizes, $A_{t + \gamma}$ can take any
value between those two limits.

\subsection{Cross section for $q \, g \rightarrow t \, Z$}
\label{sec:tz}

We can perform analysis similar to those of the previous section for
the associated production of a top and a Z boson. We computed an
analytical expression for the cross section of this process, which
is given by the sum of three terms,
\begin{equation}
\frac{d\sigma_{q g \rightarrow t Z}}{dt}\;=\;\frac{d\sigma^{EW}_{q g
\rightarrow t Z}}{dt} \;+\;\frac{d\sigma^{S}_{q g \rightarrow t
Z}}{dt} \;+\;\frac{d\sigma^{Int}_{q g \rightarrow t Z}}{dt} \;\;\;,
\label{eq:tz}
\end{equation}
with strong FCNC contributions ($\sigma^S$), electroweak ones
($\sigma^{EW}$) and interference terms between both sectors. The
expression for $d\sigma^{S}_{q g \rightarrow t Z}/dt$ was first
given in ref.~\cite{Ferr3}. The remaining formulae are quite
lengthy, involving many different combinations of anomalous
couplings with complicated coefficients. We present them in
Appendix~\ref{sec:apz} for completeness. To examine the values of
these cross sections at the LHC, we used the set of anomalous
couplings generated in the previous section, complemented with
randomly generated values for the $\eta$ and $\theta$
couplings~\footnote{Which, recall, do not contribute to FCNC
interactions involving the photon, only the Z.} and integrated the
expressions~\eqref{eq:tz} with the CTEQ6M pdf's. We chose
$\mu_F\,=\,m_t\,+\,m_Z$ and imposed a 10 GeV cut on the transverse
momentum of the particles in the final state.

\begin{figure}[htbp]
  \begin{center}
    \epsfig{file=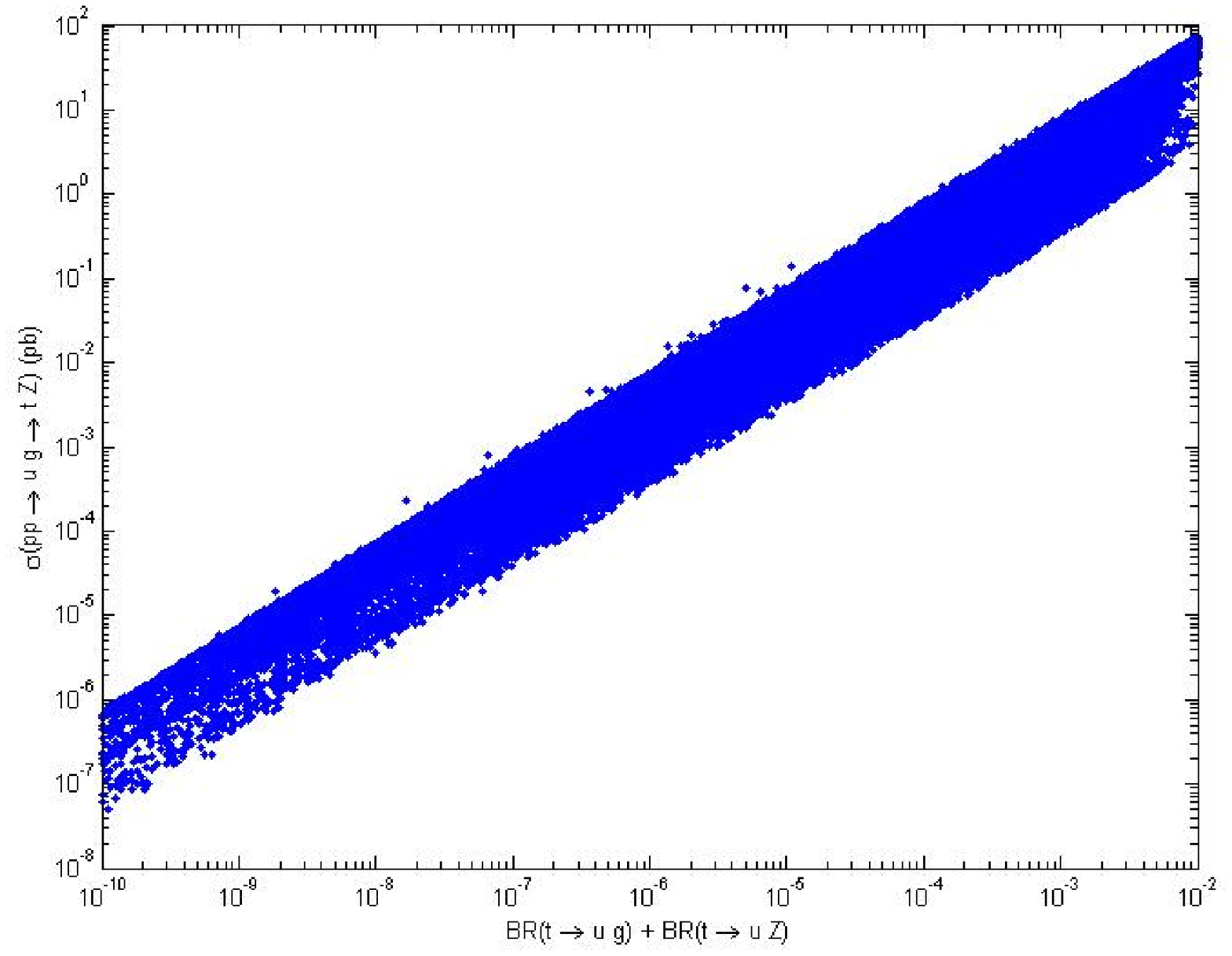,width=17 cm}
    \caption{Total (electroweak and strong contributions) cross
    section for the process $p\,p\rightarrow\,u\,g\rightarrow t\,Z$ versus
    the sum of the FCNC branching ratios for the decays
    $t\,\rightarrow\,u\,Z$ and $t\,\rightarrow\,u\,g$.}
    \label{fig:tz}
  \end{center}
\end{figure}
\begin{figure}[htbp]
  \begin{center}
    \epsfig{file=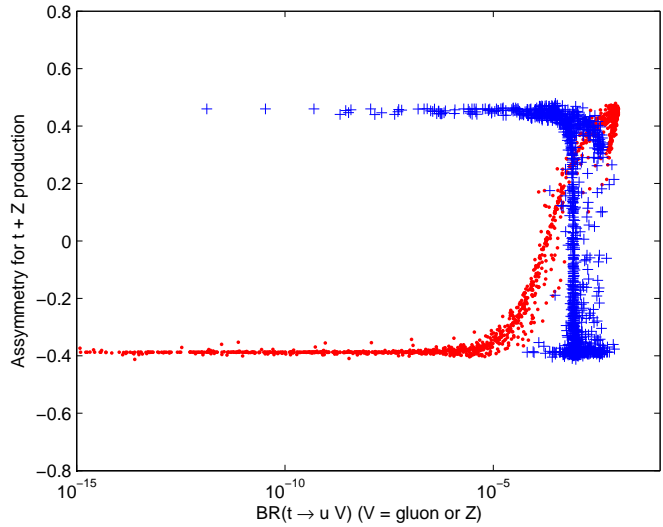,width=10 cm}
    \caption{The angular asymmetry coefficient $A_{t + Z}$
    as a function of the branching ratios $Br(t\,\rightarrow\,u\,Z)$ (crosses)
    and $Br(t\,\rightarrow\,u\,g)$ (dots).}
    \label{fig:assz}
  \end{center}
\end{figure}

Unlike what was observed for the $t\,\gamma$ channel, there is no
direct proportionality between $\sigma^{EW}(pp\rightarrow u
g\rightarrow t Z)$ and $Br(t\,\rightarrow\,q\,Z)$ - this is due to
the many different functions multiplying the several combinations of
anomalous couplings presented in Appendix~\ref{sec:apz}. Because the
functions $F_{1_Z}$ and $F_{2_Z}$ (eqs.~\eqref{eq:a12}) are very
similar, there is an {\em approximate} proportionality between the
branching ratio and $\sigma^S(pp\rightarrow u g\rightarrow t Z)$, as
was seen in ref.~\cite{Ferr3}. In fig.~\eqref{fig:tz} we plot the
total cross section for this process against the sum
$Br(t\,\rightarrow\,u\,Z)\,+\,Br(t\,\rightarrow\,u\,g)$. We see,
from this plot, that the cross section for $t\,+\,Z$ production is
always contained between two straight lines, and it is easy to
obtain the following relation, valid for the overwhelming majority
of the points shown in fig.~\eqref{fig:tz}:
\begin{equation}
200\, \left[Br(t \rightarrow u\,g)\,+\,Br(t \rightarrow
u\,Z)\right]\,<\, \sigma(p p \rightarrow u\,g \rightarrow t\,
Z)\,<\, 10^4\, \left[Br(t \rightarrow u\,g)\,+\,Br(t \rightarrow
u\,Z)\right] \,\mbox{(pb)}. \label{eq:rb}
\end{equation}
The thick band observed in this figure means any bounds obtained,
say, on the cross section, will translate into a less severe bound
on the sum of the branching ratios than what happened for the
$t\,+\,\gamma$ channel. For instance, in fig.~\eqref{fig:sigbrs} an
upper bound on the cross section $\sigma(pp\rightarrow u
g\rightarrow t \gamma)$ of $10^{-2}$ implied
$Br(t\,\rightarrow\,u\,\gamma)\,+\,Br(t\,\rightarrow\,u\,g)\,<\,
10^{-5}$, whereas a similar bound on $\sigma(pp\rightarrow u
g\rightarrow t Z)$ gives us approximately, from the right-hand side
of the band in fig.~\eqref{fig:tz},
$Br(t\,\rightarrow\,u\,Z)\,+\,Br(t\,\rightarrow\,u\,g)\,<\,
10^{-4}$. If we didn't have this band of values, but rather a line
corresponding to its left-hand side edge, the bound would be one
order of magnitude lower. As before, we obtain qualitatively
identical results for the processes involving the $c$ quark, and we
can quote rough bounds similar to those of eq.~\eqref{eq:rb},
\begin{equation}
30\, \left[Br(t \rightarrow c\,g)\,+\,Br(t \rightarrow
c\,Z)\right]\,<\, \sigma(p p \rightarrow c\,g \rightarrow t\,
Z)\,<\, 600\, \left[Br(t \rightarrow c\,g)\,+\,Br(t \rightarrow
c\,Z)\right] \,\mbox{(pb)}.
\end{equation}
\begin{figure}[htbp]
  \begin{center}
    \epsfig{file=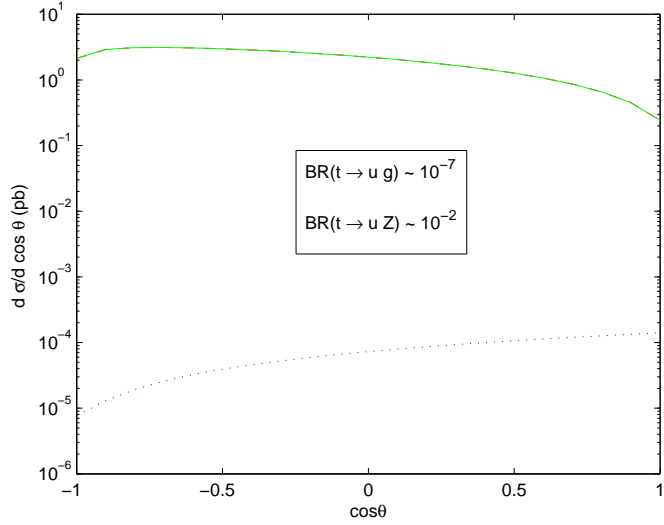,width=10 cm}
    \caption{Differential cross sections for the process $p p \rightarrow u g
\rightarrow t Z$. Total (thick line), electroweak (dashed line) and
strong (dotted line) contributions. The electroweak contribution
practically coincides with the strong one.}
    \label{fig:difzf}
  \end{center}
\end{figure}
And again, we observe that the strong and electroweak cross sections
have different angular dependencies. In fig.~\eqref{fig:difzf} we
plot the differential cross section for the process $p p \rightarrow
u g \rightarrow t Z$, both the strong and electroweak contributions,
for a typical choice of anomalous couplings for which the
electroweak FCNC interactions dominate over the strong ones. The
strong contributions increase with $\cos\theta$, whereas the
electroweak ones decrease. If the strong FCNC couplings dominate
over the electroweak ones, then the total cross section would very
closely mimic the angular dependence of the dotted line in
fig.~\eqref{fig:difzf}. Once more, if the electroweak and strong
FCNC interactions have contributions of similar magnitudes, then it
will not be possible to distinguish them through this analysis. We
can define an asymmetry coefficient for the $t\,+\,Z$ process as
well, namely
\begin{equation}
A_{t + Z} \;=\;\frac{\sigma_{t + Z}(\cos\theta > 0) \,-\,\sigma_{t +
Z}(\cos\theta < 0)}{\sigma_{t + Z}(\cos\theta > 0) \,+\,\sigma_{t +
Z}(\cos\theta < 0)}\;\;\; .
\end{equation}
We will now use the set of anomalous couplings generated to produce
fig.~\eqref{fig:assg} and plot the evolution of $A_{t + Z}$ with
both FCNC branching ratios in fig.~\eqref{fig:assz}.
Again, we see a clear distinction between dominance of electroweak
FCNC interactions or strong FCNC ones. In the former case $A_{t +
Z}$ tends to a value of approximately $0.4$, and in the latter
situation we have $A_{t + Z}\,\sim\,-0.4$ - this is particulary
interesting since the asymmetry changes signs, going from one regime
to the other. Once more, if both branching ratios have like sizes,
$A_{t + Z}$ may have any value between these two extrema.

\section{Discussion and conclusions}
\label{sec:disc}

Even if the top quark has indeed large FCNC branching ratios -
strong or electroweak ones -, which would lead to significant cross
sections of associated single top production at the LHC, could those
processes actually be observed? In other words, given the numerous
backgrounds present at the LHC, is it possible to extract a
meaningful FCNC signal from the expected data? The very thorough
analysis of ref.~\cite{del Aguila:1999ec} seems to indicate so. For
instance, for $t\,+\,Z$ production they identify several possible
channels available to identify the FCNC signal, summarised in
table~\ref{tab:tZ}.
\begin{table}[t]
\begin{center}
\begin{tabular}{cccccc}\hline\hline \\
 Final State & \hspace{2cm} & Fraction (\%) & \hspace{2cm} &
 Backgrounds
\\ & & & \\ \hline \\
$tZ \rightarrow (bjj) \, (jj)$ &  & 22.2 & & $jjjjj$ &   \vspace{0.5cm} \\
$tZ \rightarrow (bjj) \, (\nu \bar{\nu})$ &  & 8.1 & & $t \bar{t}$, $Wt$, $Zjjj$ &   \vspace{0.5cm} \\
$tZ \rightarrow (bl \nu) \, (jj)$ &  & 7.5 & & $t \bar{t}$, $Wt$, $Wjjj$ & \vspace{0.5cm} \\
$tZ \rightarrow (bl \nu) \, (\nu \bar{\nu})$ &  & 2.7 & & $Wj$ &\vspace{0.5cm} \\
$tZ \rightarrow (bjj) \, (ll)$ &  & 2.3 & & $Zjjj$, $ZWj$ &  \vspace{0.5cm} \\
$tZ \rightarrow (bjj) \, (\bar{b} b)$ &  & 2.2 & & $b \bar{b} jjj$ &\vspace{0.5cm} \\
$tZ \rightarrow (bl \nu) \, (ll)$ &  & 0.8 &   & $ZWj$ & \vspace{0.5cm} \\
$tZ \rightarrow (bl \nu) \, (\bar{b} b)$ &  & 0.7 & & $t \bar{t}$,
$Wt$, $ZWj$, $W b \bar{b} j$& \\ & & &
\\\hline\hline
\end{tabular}
\caption{Possible final states in $t \, Z$ production, and main
backgrounds to each process~\cite{del Aguila:1999ec}.}
\end{center}
\label{tab:tZ}
\end{table}
For all of these processes, the processes $W Z j$, $t\bar{t}$ and
single top production will also act as backgrounds. It is also
likely, considering the immense QCD backgrounds, that only those
processes with at least one lepton will be possible to observe at
the LHC. To build this table, the top quark was considered to decay
according to SM physics, $t \rightarrow b \, W$, and the several
decay possibilities within the SM of the $W$ and $Z$ bosons give the
possibilities listed therein. The fraction attributed to each
channel corresponds to the percentages of each decay mode of the $W$
and $Z$ as well as a 90 \% tagging efficiency for lepton (electron
or muon) tagging, and a 60 \% one for each b-jet. The most
impressive result of ref.~\cite{del Aguila:1999ec}, though, is the
efficiency with which the FCNC signal is extracted from these
backgrounds: they have shown that a battery of simple kinematical
cuts on the observed particles is more than enough to obtain a very
clear - and statistically meaningful - FCNC signal. For $t\,+\,Z$
production they conclude that the best channel would be
$p\,p\rightarrow t\,Z\rightarrow l^+\,l^-\,l\,\nu\,b$. For
$t\,+\,\gamma$ production the analysis is made simpler by the photon
not having decay branching ratios, which aides the statistics
obtained - the best channel available would be $p\,p\rightarrow
t\,\gamma\rightarrow \gamma\,l\,\nu\,b$. Clearly, only an analysis
analogous to that of~\cite{del Aguila:1999ec}, with the FCNC
interactions considered in the present paper included in an event
generator, would be capable of reaching definite conclusions
regarding which kinematical cuts would be better suited to obtain a
clear FCNC signal. That study is beyond the scope of the present
paper, though a preliminary study of our strong FCNC interactions in
the LHC environment, using the TopReX event generator~\cite{toprex},
is about to be concluded~\cite{uson}. A word on higher-order QCD
corrections: they are manifestly difficult to compute in the
effective operator formalism, since the lagrangian becomes
non-renormalizable. A recent work using electroweak top FCNC
couplings~\cite{kfac}, however, concluded that those corrections
greatly reduce any dependence the results obtained at tree level
might have on the scales of renormalization and factorization. These
authors have also shown that the higher order corrections tend to
slightly increase the leading order result.

To summarise, we employed the effective operator formalism to
parameterize the effects of any theory that might have as its
low-energy limit the SM. The fact that we are working in a gauge
invariant formalism allowed us to find many relations between
couplings and quantities which, {\em a priori}, would not be related
at all. In particular we found a near-proportionality between the
cross section of associated top plus photon production at the LHC
and the sum of the FCNC decays of the top to a photon and a gluon.
We estimated the cross sections for $t\,+\,\gamma$ and $t\,+\,Z$
production at the LHC and saw that, for large enough values of the
top FCNC branching ratios, one might expect a significant number of
events. We also concluded that, for these processes, the interplay
between the strong and electroweak anomalous interactions tends to
increase the values of the cross sections - the interference between
both FCNC sectors is mostly constructive. The analysis of the
differential cross sections for $t\,+\,\gamma$ and $t\,+\,Z$
production will possibly allow the identification of the source of
FCNC physics - the strong or the electroweak sector.

\vspace{0.25cm} {\bf Acknowledgments:} Our thanks to Nuno Castro,
Filipe Veloso and Ant\'onio Onofre for a careful reading of the
manuscript and many interesting discussions. This work is supported
by Funda\c{c}\~ao para a Ci\^encia e Tecnologia under contract
POCI/FIS/59741/2004 and PTDC/FIS/70156/2006. R.S. is supported by
FCT under contract SFRH/BPD/23427/2005. R.G.J. is supported by FCT
under contract SFRH/BD/19781/2004.

\appendix

\section{Cross section expression for the process $q \, g \rightarrow t \, Z$}
\label{sec:apz}

As mentioned in section~\ref{sec:tz} the cross section for the
associated production of a top and a Z boson is given by three
terms, as in eq.~\eqref{eq:tz}. The strong FCNC contribution is
given by:
\begin{eqnarray}
\frac{d\, \sigma^S_{q \, g \rightarrow t \, Z}}{dt} & = &
\frac{e^2}
  {96 \, \pi \, s^2 \, \Lambda^4}
\left[ F_{1_Z}(t,s) \, \left\{ \left| \alpha^S_{qt} +
(\alpha^S_{tq})^* \right|^2 + \frac{8v}{m_t} \, Im \left[
(\alpha^S_{qt} + ( \alpha^S_{tq})^*) \, \beta^S_{tq} \right] +
\frac{16v^2}{m_t^2} \, \left| \beta^S_{tq} \right|^2 \right\} +
\right. \nonumber \\[0.25cm]
&& \qquad \qquad \left. + \, F_{2_Z}(t,s) \,  \frac{16v^2}{m_t^2} \,
\left| \beta^S_{qt} \right|^2 \right]\;\;\;,
\end{eqnarray}
with coefficients
\begin{eqnarray}
F_{1_Z}(t,s) & = & \frac{- m_t^2}{72\,
    c_W^2\,m_Z^2\,{\left( m_t^2 - s \right) }^2 \,s_W^2\,t^2}
    \left[
32\,m_t^8\,m_Z^2\,
         s_W^4\,\left( m_Z^2 - t \right)  +
        32\,m_t^4\,m_Z^2\,s_W^4\,
         \left( m_Z^2 - t \right) \,\left( s^2 + 4\,s\,t + t^2 \right)
         + \right. \nonumber \\[0.25cm]
&&        \qquad \qquad s^2\,t^2\,\left( 2\,m_Z^4\,
            \left( 9 - 24\,s_W^2 + 32\,s_W^4 \right)  + 9\,s\,t -
           2\,m_Z^2\,\left( 9 - 24\,s_W^2 + 32\,s_W^4 \right)
              \,\left( s + t \right)  \right) \, +    \nonumber \\[0.25cm]
&&        \qquad \qquad
         m_t^2\,s\,t\,\left( -9\,s\,t^2 -64\,m_Z^4\,s_W^4\,\left( s + t \right) +
           m_Z^2\,\left( 32\,s^2\,s_W^4 +
              3\,s\,\left( 3 - 32\,s_W^2 + 64\,s_W^4 \right) \,t +
              96\,s_W^4\,t^2 \right)  \right) - \nonumber \\[0.25cm]
&&        \qquad \qquad \left. 32\,m_t^6\,m_Z^2\,s_W^4\,
         \left( 2\,m_Z^2\,\left( s + t \right)-t\,\left( s + 2\,t \right)
         \right)\right]
 \nonumber  \\[0.35cm]
F_{2_Z}(t,s) & = & \frac{ m_t^2}{72 \,
    c_W^2\,m_Z^2\,{\left( m_t^2 - s \right) }^2 \,s_W^2\,t^2}
    \left[
- 2\,m_t^4\,m_Z^2\,{\left( 3 - 4\,s_W^2 \right) }^2\,
   \left( m_Z^2 - t \right) \,\left( s^2 + 4\,s\,t + t^2 \right)
         + \right. \nonumber \\[0.25cm]
&&        \qquad \qquad s^2\,t^2\,\left( -2\,m_Z^4\,
      \left( 9 - 24\,s_W^2 + 32\,s_W^4 \right)  - 9\,s\,t +
     2\,m_Z^2\,\left( 9 - 24\,s_W^2 + 32\,s_W^4 \right) \,
      \left( s + t \right)  \right)     +
        \nonumber \\[0.25cm]
&&        \qquad \qquad m_t^2\,s\,t\,\left\{ 9\,s\,t^2 +
     4\,m_Z^4\,{\left( 3 - 4\,s_W^2 \right) }^2\,\left( s + t \right)  +
     m_Z^2\,\left( -2\,s^2\,{\left( 3 - 4\,s_W^2 \right) }^2
     \right. - \right.
\nonumber \\[0.25cm]
&&         \qquad \qquad \left. \left. 3\,s\,\left( 15 + 64\,\left(
-s_W^2 + s_W^4 \right) \right) \,t -
        6\,{\left( 3 - 4\,s_W^2 \right) }^2\,t^2 \right)
        \right\}
\nonumber \\[0.25cm]
&&        \qquad \qquad \left. 2\,m_t^8\,m_Z^2\,{\left( 3 - 4\,s_W^2
\right) }^2\, \left( -m_Z^2 + t \right) + 2\,m_t^6\,m_Z^2\,
   {\left( 3 - 4\,s_W^2 \right) }^2\,
   \left( 2\,m_Z^2\,\left( s + t \right)- t\,\left( s + 2\,t \right)  \right)
   \right]\;\;\; .
   \label{eq:a12}
\end{eqnarray}
The electroweak FCNC contribution is given by the following
expression:
\begin{eqnarray}
\frac{d \, \sigma^{EW}_{q \, g \rightarrow t \, Z}}{dt} & = &
\frac{g_s^2}
  {96 \, \pi \, s^2  \, \Lambda^4}
\left[ G_{1_Z}(t,s) \, \left| \alpha^Z_{qt} \right|^2 + G_{2_Z}(t,s)
\, \left| \alpha^Z_{tq} \right|^2 + G_{3_Z}(t,s) \, ( \left|
\beta^Z_{qt} \right|^2 + \left| \beta^Z_{tq} \right|^2)+
G_{4_Z}(t,s) \, ( \left| \eta_{qt} \right|^2 + \left|
\bar{\eta}_{qt} \right|^2) \right.
\nonumber \\[0.25cm]
&&  \qquad + \, G_{5_Z}(t,s) \, \left| \theta \right|^2 +
G_{6_Z}(t,s) \, Re \left[ \alpha^Z_{qt} \, \alpha^Z_{tq} \right] +
G_{7_Z}(t,s) \, Im \left[ \alpha^Z_{qt} \, \beta^Z_{tq} \right]
\nonumber \\[0.25cm]
&&  \qquad + \, G_{8_Z}(t,s) \, Im \left[ \alpha^{Z^*}_{tq} \,
\beta^Z_{tq} \right] + G_{9_Z}(t,s) \, Re \left[ \alpha^Z_{qt}
\theta^* \right]+G_{10_Z}(t,s) \, Re \left[ \alpha^Z_{tq} \theta
\right]
\nonumber \\[0.25cm]
&& \qquad   \left. + \, G_{11_Z}(t,s) \, Re \left[ \beta^Z_{qt}
(\eta_{qt}-\bar{\eta}_{qt})^* \right] + G_{12_Z}(t,s) \, Im \left[
\beta^Z_{tq} \, \theta \right] + G_{13_Z}(t,s) \, Re \left[
\eta_{qt} \bar{\eta}_{qt}^* \right] \right]
\end{eqnarray}
where the $G_{i_Z}$ functions are given by
\begin{eqnarray}
G_{1_Z}(t,s) & = & \frac{1}{4\,s\,{\left( m_t^2 - t \right) }^2}
\left[ m_t^{10} + m_t^8\,\left( 2\,m_Z^2 - 2\,s - t \right) +
    m_t^6\,\left( -5\,m_Z^4 + s^2 + 4\,s\,t + t^2 -
       4\,m_Z^2\,\left( s + t \right)  \right)  \right.\nonumber \\[0.25cm]
 &&     +
    m_t^2\,\left( 2\,m_Z^8 - 8\,m_Z^6\,t -
       4\,m_Z^2\,t^2\,\left( s + t \right)  + 2\,s\,t^2\,\left( s + t \right)  +
       m_Z^4\,{\left( s + 3\,t \right) }^2 \right) \nonumber \\[0.25cm]
 &&  \left. - m_Z^4\,t\,\left( 2\,m_Z^4 + s^2 + t^2 -
       2\,m_Z^2\,\left( s + t \right)  \right)  +
    m_t^4\,\left( 6\,m_Z^6 - 3\,m_Z^4\,t +
       2\,m_Z^2\,t\,\left( s + 3\,t \right)  - t\,\left( 3\,s^2 + 6\,s\,t + t^2 \right)
       \right) \right]
 \nonumber  \\[0.3cm]
G_{2_Z}(t,s) & = & \frac{1}{4\,s\,
    {\left( m_t^2 - t \right) }^2}
\left[
    m_t^{10} - m_t^8\,\left( 4\,m_Z^2 +
2\,s + t \right)  +
    m_t^6\,\left( 7\,m_Z^4 + s^2 + 2\,m_Z^2\,t + 4\,s\,t +
       t^2 \right) \right. \nonumber \\[0.25cm]
 &&          - m_Z^4\,t\,
     \left( 2\,m_Z^4 + s^2 + t^2 - 2\,m_Z^2\,\left( s + t \right)  \right)
     - m_t^4\,\left( 6\,m_Z^6 + 3\,m_Z^4\,t -
       2\,m_Z^2\,s\,\left( s + 3\,t \right)  + t\,\left( 3\,s^2 + 6\,s\,t + t^2 \right)
       \right)\nonumber \\[0.25cm]
 &&    \left. + m_t^2\,\left( 2\,m_Z^8 + 4\,m_Z^6\,t +
       2\,s\,t^2\,\left( s + t \right)  +
       m_Z^4\,\left( s^2 - 6\,s\,t - 3\,t^2 \right)  +
       m_Z^2\,\left( -2\,s^2\,t + 2\,t^3 \right)  \right) \right]
\nonumber \\[0.3cm]
G_{3_Z}(t,s) & = & \frac{2\, v^2} {s\,
    {\left( m_t^2 - t \right) }^2}  \left[ 2\,m_t^8 -
      m_t^6\,\left( 3\,m_Z^2 + 4\,s + 2\,t \right)  -
      t\,\left( 2\,m_Z^6 - 2\,m_Z^4\,\left( s + t \right)  -
         4\,s\,t\,\left( s + t \right)  + m_Z^2\,{\left( s + t \right) }^2 \right)  \right. \nonumber \\[0.25cm]
 &&          +
      m_t^4\,\left( 2\,m_Z^4 - m_Z^2\,\left( 2\,s + t \right)  + 2\,\left( s^2 + 4\,s\,t + t^2 \right)
         \right)   \nonumber \\[0.25cm]
 &&
         \left.  + m_t^2\,\left( 2\,m_Z^6 - 4\,m_Z^4\,t -
         2\,t\,\left( 3\,s^2 + 6\,s\,t + t^2 \right)+
         m_Z^2\,\left( s^2 + 6\,s\,t + 5\,t^2 \right)  \right)
         \right] \nonumber
\end{eqnarray}
\begin{eqnarray}
G_{4_Z}(t,s) & = & \frac{v^2}{8\,
    m_Z^2\,s\,{\left( m_t^2 - t \right) }^2} \left[  m_t^{10} - m_t^8\,
       \left( 4\,m_Z^2 + 2\,s + t \right)  +
      m_t^6\,\left( 7\,m_Z^4 + 2\,m_Z^2\,t +
         {\left( s + t \right) }^2 \right)  \right. \nonumber \\[0.25cm]
 &&
      - m_Z^2\,t\,\left( 2\,m_Z^6 -
         2\,m_Z^4\,\left( s + t \right)  - 4\,s\,t\,\left( s + t \right)  +
         m_Z^2\,{\left( s + t \right) }^2 \right) \nonumber \\[0.25cm]
 &&  - m_t^4\,\left( 6\,m_Z^6 + t\,{\left( s + t \right) }^2 +
         m_Z^4\,\left( 2\,s + 3\,t \right)  - 2\,m_Z^2\,s\,\left( 3\,s + 5\,t \right) \right) \nonumber \\[0.25cm]
 &&  \left. + m_t^2\,m_Z^2\,
       \left( 2\,m_Z^6 + 4\,m_Z^4\,t +
         m_Z^2\,\left( s^2 - 2\,s\,t - 3\,t^2 \right)  +
         2\,t\,\left( -5\,s^2 - 4\,s\,t + t^2 \right)  \right)
         \right] \nonumber  \\[0.3cm]
G_{5_Z}(t,s) & = & \frac{v^4} {2\,m_Z^2\,s\,
    {\left( m_t^2 - t \right) }^2}  \left[ m_t^8 - m_t^6\,\left(
2\,s + t \right)  -
      2\,m_Z^2\,t\,\left( 2\,m_Z^4 + s^2 + t^2 -
         2\,m_Z^2\,\left( s + t \right)  \right) \right. \nonumber \\[0.25cm]
&&  \left. +
      m_t^4\,\left( -2\,m_Z^4 - 2\,m_Z^2\,t +
         {\left( s + t \right) }^2 \right)  +
      m_t^2\,\left( 4\,m_Z^6 - 2\,m_Z^4\,t -
         t\,{\left( s + t \right) }^2 + 2\,m_Z^2\,\left( s^2 - s\,t + 2\,t^2 \right)
         \right) \right] \nonumber \\[0.3cm]
G_{6_Z}(t,s) & = & \frac{1}{2\,s\,{\left( m_t^2 - t \right) }^2}
\left[ m_t^{10} - m_t^8\,\left( 2\,m_Z^2 + 2\,s + t \right) +
    m_t^6\,\left( m_Z^4 + s^2 + 4\,s\,t + t^2 \right) \right.
\nonumber \\[0.25cm]
&&
     +
    m_Z^4\,t\,\left( 2\,m_Z^4 + s^2 + t^2 -
       2\,m_Z^2\,\left( s + t \right)  \right) -
    m_t^2\,\left( 2\,m_Z^8 +
       m_Z^4\,{\left( s - t \right) }^2 - 2\,s\,t^2\,\left( s + t \right)  \right)
\nonumber \\[0.25cm]
&&     \left.     +
    m_t^4\,\left( 2\,m_Z^6 - m_Z^4\,t +
       2\,m_Z^2\,t\,\left( s + t \right)  - t\,\left( 3\,s^2 + 6\,s\,t + t^2 \right)
       \right) \right]\nonumber
\end{eqnarray}
\begin{eqnarray}
G_{7_Z}(t,s) & = & \frac{2 \,m_t \,v}{s\,
    {\left( m_t^2 - t \right) }^2} \left[ m_t^8 - m_t^6\,\left(
2\,s + t \right) +
      m_t^4\,\left( -2\,m_Z^4 + s^2 + 4\,s\,t + t^2 -
         2\,m_Z^2\,\left( s + t \right)  \right) \right.
\nonumber \\[0.25cm]
&&        +
      2\,t\,\left( -2\,m_Z^6 + 2\,m_Z^4\,\left( s + t \right)  -
         m_Z^2\,t\,\left( s + t \right) + s\,t\,\left( s + t \right)  \right)
\nonumber \\[0.25cm]
&&        \left.    +
      m_t^2\,\left( 4\,m_Z^6 - 2\,m_Z^4\,t +
         2\,m_Z^2\,t\,\left( s + 2\,t \right)  -
         t\,\left( 3\,s^2 + 6\,s\,t + t^2 \right)  \right) \right]
         \nonumber \\[0.3cm]
G_{8_Z}(t,s) & = & \frac{2 \,m_t \,v}{s\,
    {\left( m_t^2 - t \right) }^2}
\left[ m_t^8 -
      m_t^6\,\left( 3\,m_Z^2 + 2\,s + t \right)  +
      m_t^4\,\left( 4\,m_Z^4 + s^2 + m_Z^2\,t + 4\,s\,t +
         t^2 \right) \right.
\nonumber \\[0.25cm]
&&          + t\,\left( 2\,m_Z^6 -
         2\,m_Z^4\,\left( s + t \right)  + 2\,s\,t\,\left( s + t \right)  +
         m_Z^2\,\left( -s^2 + t^2 \right)  \right)
\nonumber \\[0.25cm]
&&         \left.  -
      m_t^2\,\left( 2\,m_Z^6 + 2\,m_Z^4\,t -
         m_Z^2\,\left( s^2 + 4\,s\,t + t^2 \right)  +
         t\,\left( 3\,s^2 + 6\,s\,t + t^2 \right)  \right)
         \right]
 \nonumber \\[0.3cm]
G_{9_Z}(t,s) & = & \frac{v^2}{s\,{\left( m_t^2 - t \right) }^2}
\left[ -2\,m_t^8 + m_t^4\,
       \left( -2\,m_Z^4 + m_Z^2\,t - 2\,t^2 \right) \right.
\nonumber \\[0.25cm]
&&  +   m_t^6\,\left( 3\,m_Z^2 + 2\,\left( s + t \right) \right) +
      m_Z^2\,t\,\left( 2\,m_Z^4 + s^2 + t^2 -
         2\,m_Z^2\,\left( s + t \right)  \right)
\nonumber \\[0.25cm]
&&      \left. -
      m_t^2\,\left( 2\,m_Z^6 - 4\,m_Z^4\,t -
         2\,t^2\,\left( s + t \right)  + m_Z^2\,\left( s^2 + 2\,s\,t + 5\,t^2 \right)
         \right)  \right]
\end{eqnarray}
\begin{eqnarray}
G_{10_Z}(t,s) & = &  \frac{-v^2}{s\,
      {\left( m_t^2 - t \right) }^2} \left[ m_t^8 -
        m_t^6\,\left( 3\,m_Z^2 + t \right)  +
        m_t^4\,\left( 4\,m_Z^4 - s^2 + m_Z^2\,t - 2\,s\,t +
           t^2 \right) \right.
\nonumber \\[0.25cm]
&&            \left. + m_Z^2\,t\,
         \left( 2\,m_Z^4 + s^2 + t^2 - 2\,m_Z^2\,\left( s + t \right)
           \right)  - m_t^2\,
         \left( 2\,m_Z^6 + 2\,m_Z^4\,t - s^2\,t + t^3 +
           m_Z^2\,\left( s^2 - 4\,s\,t - t^2 \right)  \right)
           \right]
         \nonumber \\[0.3cm]
G_{11_Z}(t,s) & = & \frac{ v^2}{s\,{\left( m_t^2 - t \right) }^2}
\left[ m_t^8 - m_t^6\,
       \left( 3\,m_Z^2 - 2\,s + t \right)  +
      t\,\left( 2\,m_Z^6 - 2\,m_Z^4\,\left( s + t \right)  -
         4\,s\,t\,\left( s + t \right) + m_Z^2\,{\left( s + t \right) }^2 \right)  \right.
\nonumber \\[0.25cm]
&&           +
      m_t^4\,\left( 4\,m_Z^4 - 3\,s^2 - 10\,s\,t + t^2 +
         m_Z^2\,\left( 2\,s + t \right)  \right)
\nonumber \\[0.25cm]
&&         \left.      -
      m_t^2\,\left( 2\,m_Z^6 + 2\,m_Z^4\,t +
         m_Z^2\,\left( s^2 - t^2 \right)  + t\,\left( -7\,s^2 - 10\,s\,t + t^2 \right)
         \right)  \right]
 \nonumber \\[0.3cm]
G_{12_Z}(t,s) & = & \frac{-2\, m_t\, v^3}{s\,
    {\left( m_t^2 - t \right) }^2} \,\left[ 3\,m_t^6 -
      m_t^4\,\left( 6\,m_Z^2 + 2\,s + 3\,t \right)  +
      m_t^2\,\left( 6\,m_Z^4 - s^2 - 2\,s\,t + 3\,t^2 \right)
      \right.
\nonumber \\[0.25cm]
&&       +
      \left. t\,\left( -6\,m_Z^4 + s^2 - 2\,s\,t - 3\,t^2 +
         6\,m_Z^2\,\left( s + t \right)  \right)  \right]
          \nonumber \\[0.3cm]
G_{13_Z}(t,s) & = & \frac{- v^2} {4\,
    m_Z^2\,s\,{\left( m_t^2 - t \right) }^2}  \left[ m_t^{10} -
        m_t^8\,\left( 4\,m_Z^2 + 2\,s + t \right)  +
        m_t^6\,\left( 7\,m_Z^4 + 2\,m_Z^2\,t +
           {\left( s + t \right) }^2 \right) \right.
\nonumber \\[0.25cm]
&&            -
        m_Z^2\,t\,\left( 2\,m_Z^6 -
           2\,m_Z^4\,\left( s + t \right)  - 4\,s\,t\,\left( s + t \right)  +
           m_Z^2\,{\left( s + t \right) }^2 \right)
\nonumber \\[0.25cm]
&&           -
        m_t^4\,\left( 6\,m_Z^6 + t\,{\left( s + t \right) }^2 +
           m_Z^4\,\left( 2\,s + 3\,t \right)  -
           2\,m_Z^2\,s\,\left( 3\,s + 5\,t \right)  \right)
\nonumber \\[0.25cm]
&&        \left.    +
        m_t^2\,m_Z^2\,
         \left( 2\,m_Z^6 + 4\,m_Z^4\,t +
           m_Z^2\,\left( s^2 - 2\,s\,t - 3\,t^2 \right)  +
           2\,t\,\left( -5\,s^2 - 4\,s\,t + t^2 \right)  \right)
           \right]\;\;\; .
\end{eqnarray}
Finally, the strong-electroweak interference cross section is given
by
\begin{eqnarray}
\frac{d \, \sigma^{Int}_{q \, g \rightarrow t \, Z}}{dt} & = &
\frac{e \, g_s}
  {96 \, \pi \, s^2  \, \Lambda^4}
\left[ H_{1_Z}(t,s) \, \left\{ Re \left[ \left( \alpha^S_{qt} +
\alpha^{S^*}_{tq} \right) \alpha^{Z^*}_{qt} \right] + \frac{4 \,
v}{mt} \, Im \left[ \beta^S_{tq} \, \alpha^Z_{qt} \right] \right\}
\right.
\nonumber \\[0.25cm]
&&  \qquad + H_{2_Z}(t,s) \, \left\{ Re \left[ \left( \alpha^S_{qt}
+ \alpha^{S^*}_{tq} \right) \alpha^{Z}_{tq} \right] + \frac{4 \,
v}{mt} \, Im \left[ \beta^S_{tq} \, \alpha^{Z^*}_{tq} \right]
\right\}
\nonumber \\[0.25cm]
&&  \qquad + H_{3_Z}(t,s) \, \left\{ Im \left[ \left( \alpha^S_{qt}
+ \alpha^{S^*}_{tq} \right) \beta^{Z}_{tq} \right] + \frac{4 \,
v}{mt} \, Re \left[ \beta^{S^*}_{tq} \, \beta^{Z}_{tq} \right]
\right\}
\nonumber \\[0.25cm]
&&  \qquad + H_{4_Z}(t,s) \, \left\{ Re \left[ \left( \alpha^S_{qt}
+ \alpha^{S^*}_{tq} \right) \theta^* \right] + \frac{4 \, v}{mt} \,
Im \left[ \beta^{S}_{tq} \, \theta \right] \right\}+ H_{5_Z}(t,s) \,
Re \left[  \beta^S_{qt} \, \beta^{Z^*}_{qt} \right]
\nonumber \\[0.25cm]
&& \qquad   \left. + H_{6_Z}(t,s) \,  Re \left[  \beta^S_{qt} \,
\left( \eta_{qt} - \bar{\eta}_{qt} \right)^* \right] \right]
\end{eqnarray}
with $H_{i_Z}$ given by
\begin{eqnarray}
H_{1_Z}(t,s) & = & \frac{m_t^2}{6\,c_W\,\left( m_t^2 - s \right) \,
    s_W\,\left( m_t^2 - t \right) \,t}
\left[ 12\,m_t^6\,s_W^2\,t    -
      m_t^4\,\left( 4\,m_Z^4\,s_W^2 +
         t\,\left( -3\,s + 16\,s\,s_W^2 + 16\,s_W^2\,t \right)  \right) \right.
\nonumber \\[0.25cm]
&&       + t\,\left( 4\,m_Z^4\,s_W^2\,\left( -s + t \right)  +
         s\,t\,\left( 3\,s - 4\,s\,s_W^2 - 4\,s_W^2\,t \right)  +
         2\,m_Z^2\,s\,\left( 2\,s\,s_W^2 - 3\,t +
            4\,s_W^2\,t \right)  \right)
\nonumber \\[0.25cm]
&&     \left.  +
      m_t^2\,\left( 4\,m_Z^4\,s\,s_W^2 +
         2\,m_Z^2\,s\,\left( 3 - 8\,s_W^2 \right) \,t +
         t\,\left( s^2\,\left( -3 + 4\,s_W^2 \right)  +
            3\,s\,\left( -1 + 4\,s_W^2 \right) \,t + 4\,s_W^2\,t^2 \right)
         \right)  \right]
 \nonumber  \\[0.3cm]
H_{2_Z}(t,s) & = & \frac{- m_t^2}{6\,c_W\,
    \left( m_t^2 - s \right) \,s_W\,\left( m_t^2 - t \right) \,t}
   \left[   t\,\left( m_Z^2\,s\,\left( 4\,s\,s_W^2 - 3\,t \right)  +
           4\,m_Z^4\,s_W^2\,\left( -s + t \right)  +
           s\,t\,\left( -3\,s + 4\,s\,s_W^2 + 4\,s_W^2\,t \right)
           \right)\right.
\nonumber \\[0.25cm]
&&         + 4\,m_t^6\,s_W^2\,
         \left( 2\,m_Z^2 - 3\,t \right)  + m_t^4\,\left( -4\,m_Z^4\,s_W^2 -
           8\,m_Z^2\,s\,s_W^2 +
           t\,\left( -3\,s + 16\,s\,s_W^2 + 16\,s_W^2\,t \right)  \right)
\nonumber \\[0.25cm]
&&         \left. + m_t^2\,\left( 4\,m_Z^4\,s\,s_W^2 +
           m_Z^2\,t\,\left( 3\,s - 8\,s_W^2\,t \right)  +
           t\,\left( s^2\,\left( 3 - 4\,s_W^2 \right)  +
              3\,s\,\left( 1 - 4\,s_W^2 \right) \,t - 4\,s_W^2\,t^2 \right)
               \right)  \right]
\nonumber \\[0.3cm]
H_{3_Z}(t,s) & = & \frac{m_t \, v}{3\,c_W\,\left( m_t^2 - s \right)
\,
    s_W\,\left( m_t^2 - t \right) \,t}
\left[ -8\,m_t^6\,s_W^2\,
       \left( m_Z^2 - 3\,t \right)  +
      2\,m_t^4\,\left( 4\,m_Z^2\,s\,s_W^2 +
         t\,\left( 3\,s - 16\,s\,s_W^2 - 16\,s_W^2\,t \right)
         \right)\right.
\nonumber \\[0.25cm]
&&          +
      s\,t^2\,\left( m_Z^2\,\left( -3 + 8\,s_W^2 \right)  -
         2\,\left( -3\,s + 4\,s\,s_W^2 + 4\,s_W^2\,t \right)  \right)
\nonumber \\[0.25cm]
&& \left.         +
      m_t^2\,t\,\left( m_Z^2\,
          \left( s\,\left( 3 - 16\,s_W^2 \right)  + 8\,s_W^2\,t \right)  +
         2\,\left( s^2\,\left( -3 + 4\,s_W^2 \right)  +
            3\,s\,\left( -1 + 4\,s_W^2 \right) \,t + 4\,s_W^2\,t^2 \right)
         \right)  \right]  \nonumber
\end{eqnarray}
\begin{eqnarray}
H_{4_Z}(t,s) & = & \frac{- m_t^2 \, v^2}{6\,c_W\,\left( m_t^2 - s
\right) \,
    s_W \,\left( m_t^2 - t \right) \,t}
\left[ 8\,m_t^6\,s_W^2 -
        8\,m_t^4\,\left( m_Z^2 + s \right) \,s_W^2 \right.
\nonumber \\[0.25cm]
&& \left.       +
        m_t^2\,\left( 8\,m_Z^2\,s\,s_W^2 +
           t\,\left( 9\,s - 16\,s\,s_W^2 - 8\,s_W^2\,t \right)  \right)  +
        t\,\left( -8\,m_Z^2\,s_W^2\,\left( s - t \right)  +
           s\,\left( 8\,s\,s_W^2 - 9\,t + 8\,s_W^2\,t \right)  \right)
        \right]
 \nonumber  \\[0.3cm]
H_{5_Z}(t,s) & = & \frac{4\, v^2}{3\,
    c_W \,\left( m_t^2 - s \right) \, s_W\,
    \left( m_t^2 - t \right) \,t} \left[
      s\,t^2\,\left( m_Z^2\,\left( -3 + 8\,s_W^2 \right)  -
         2\,\left( 4\,s\,s_W^2 - 3\,t + 4\,s_W^2\,t \right)
         \right)\right.
\nonumber \\[0.25cm]
&&          -2\,m_t^6\,\left( -3 + 4\,s_W^2 \right) \,
       \left( m_Z^2 - 3\,t \right) +
      2\,m_t^4\,\left( m_Z^2\,s\,
          \left( -3 + 4\,s_W^2 \right)  +
         t\,\left( s\,\left( 9 - 16\,s_W^2 \right)  +
            4\,\left( 3 - 4\,s_W^2 \right) \,t \right)  \right)
\nonumber \\[0.25cm]
&&     \left.       +
      m_t^2\,t\,\left( m_Z^2\,
          \left( s\,\left( 9 - 16\,s_W^2 \right)  +
            2\,\left( -3 + 4\,s_W^2 \right) \,t \right)  +
         2\,\left( 4\,s^2\,s_W^2 +
            6\,s\,\left( -1 + 2\,s_W^2 \right) \,t +
            \left( -3 + 4\,s_W^2 \right) \,t^2 \right)  \right)  \right]
\nonumber \\[0.3cm]
H_{6_Z}(t,s) & = & \frac{v^2}{3\,c_W\,\left( m_t^2 - s \right) \,
    s_W\,\left( m_t^2 - t \right) \,t} \left[
  -
      2\,m_t^6\,\left( -3 + 4\,s_W^2 \right) \,\left( s + 3\,t \right)  +
      m_t^4\,t\,\left( s\,\left( -15 + 16\,s_W^2 \right)  +
         6\,\left( -3 + 4\,s_W^2 \right) \,t \right) \right.
\nonumber \\[0.25cm]
&&          +
      m_t^2\,t\,\left( -6\,s^2 +
         m_Z^2\,s\,\left( -3 + 8\,s_W^2 \right)  +
         s\,\left( 15 - 16\,s_W^2 \right) \,t +
         2\,\left( 3 - 4\,s_W^2 \right) \,t^2 \right)
\nonumber \\[0.25cm]
&&     \left.      + 2\,m_t^8\,\left( -3 + 4\,s_W^2 \right)+
      s\,t^2\,\left( m_Z^2\,\left( 3 - 8\,s_W^2 \right)  +
         2\,\left( 4\,s\,s_W^2 + \left( -3 + 4\,s_W^2 \right) \,t \right)
         \right) \right]\;\;\; .
\end{eqnarray}

\end{document}